\documentclass[12pt,a4paper]{iopart}
\usepackage{iopams}
\usepackage{graphicx}
\usepackage{epsfig}
\usepackage{enumerate}
\usepackage{color}
\usepackage[subnum]{cases}
\usepackage{relsize}
\usepackage{cite}
\usepackage{tensor}
\usepackage{footmisc}
\usepackage{mathrsfs}
\usepackage{tikz}
\usepackage{multirow}
\usepackage{booktabs}
\newcommand{\CB}{{\mathcal B}}
\newcommand{\CQ}{{\mathcal Q}}

\newcommand{\CW}{{\mathcal W}}

\newcommand{\CM}{{\mathcal M}}

\newcommand{\cstEdS}{a_{\mathrm{E}}}
\newcommand{\cstLCDM}{a_\Lambda}
\newcommand{\uEdS}{{u^{\mathrm{E}}_\CM}}
\newcommand{\UEdS}{{U^{\mathrm{E}}_\CM}}

\newcommand{\psiEdS}{{\psi^{\mathrm{E}}_\CM}}
\newcommand{\uLCDM}{{u^{\Lambda}_\CM}}

\newcommand{\psiLCDM}{{\psi^{\Lambda}_\CM}}
\newcommand{\mEdS}{{m^\mathrm{E}_{\rm eff}}}
\newcommand{\mLCDM}{{m^\Lambda_{\rm eff}}}

\newcommand{\fnX}{X_\gamma}
\newcommand{\fnG}{\Gamma_\gamma}

\newcommand{\rd}{{\mathrm d}}

\newcommand{\CD}{\mathcal{D}}

\newcommand{\CR}{\mathcal{R}}

\newcommand{\average}[2]{\left\langle #1 \right\rangle_{\cal #2}}
\newcommand{\laverage}[2]{\left\langle #1 \right\rangle_{\mathcal{#2}_{\rm \bf i}}}
\font\bigastfont=cmr10 scaled \magstep 1
\newcommand{\fdot}{\hbox{\bigastfont .}}
\newcommand{\initial}[1]{{#1_{\rm \bf i}}}

\newcommand{\inI}{{\rm I}}
\newcommand{\inII}{{\rm II}}
\newcommand{\inIII}{{\rm III}}

\usepackage{hyperref}
{
 \definecolor{BLACK}{gray}{0}
 \definecolor{WHITE}{gray}{1}
 \definecolor{RED}{rgb}{1,0,0}
 \definecolor{GREEN}{rgb}{0,1,0}
\definecolor{dgreen}{rgb}{.1,.6,.1}
\definecolor{BLUE}{rgb}{0,0,1}
 \definecolor{CYAN}{cmyk}{1,0,0,0}
 \definecolor{MAGENTA}{cmyk}{0,1,0,0}
 \definecolor{YELLOW}{cmyk}{0,0,1,0}
 \definecolor{aw}{rgb}{0.2,0.5,0.75}
  }
  \hypersetup{
    colorlinks,%
    citecolor=blue,%
    filecolor=blue,%
    linkcolor=magenta,%
    urlcolor=blue
}

\definecolor{MyDarkRed}{rgb}{0.7,0,0}

\definecolor{MyGreen}{rgb}{0.0,.7,0.0}

\makeatletter
\long\def\@makefntext#1{\parindent 1em\noindent 
 \makebox[1em][l]{\footnotesize\rm$\m@th{^{\arabic{footnote}}}$}%
 \footnotesize\rm #1}
\def\@makefnmark{\hbox{$^{\arabic{footnote}}\m@th$}}
\def\@thefnmark{\arabic{footnote}}
\makeatother
\begin{document}

\date{\today}

\title[Dark Matter from Backreaction?]{Dark Matter from Backreaction? \\Collapse models on galaxy cluster scales}

\author{Quentin Vigneron and Thomas Buchert}

\address{Univ Lyon, Ens de Lyon, Univ Lyon1, CNRS, Centre de Recherche Astrophysique de Lyon UMR5574, F--69007, Lyon, France \\
\medskip
Emails: quentin.vigneron@ens--lyon.fr and buchert@ens--lyon.fr}

\begin{abstract}
In inhomogeneous cosmology, restricting attention to an irrotational dust matter model, backreaction arises in terms of the deviation of the averaged spatial scalar curvature from a constant-curvature model on some averaging domain $\CD$, $\CW_\CD$, and the kinematical backreaction $\CQ_\CD$. These backreaction variables can be modeled as an effective scalar field, called the `morphon field'. The general cosmological equations still need a closure condition to be solved. A simple example is the class of scaling solutions where $\CW_\CD$ and $\CQ_\CD$ are assumed to follow a power law of the volume scale factor $a_\CD$. But while they can describe models of quintessence, these and other models still assume the existence of dark matter in addition to the known sources. Going beyond scaling solutions by using a model for structure formation that we argue is reasonably generic, we investigate the correspondence between the morphon field and fundamental scalar field dark matter models, in order to describe dark matter as an effective phenomenon arising from kinematical backreaction and the averaged spatial curvature of the inhomogeneous Universe. While we find significant differences with those fundamental models, our main result is that the energy budget on typical collapsing domains is provided by curvature and matter in equal parts already around the turn-around time, leading to curvature dominance thereafter and increasing to a curvature contribution of $3/4$ of the energy budget at the onset of virialization. Kinematical backreaction is subdominant at early stages, but its importance rises quickly after turn-around and dominates the curvature contribution in the final phase of the collapse. We conclude that backreaction can indeed mimic dark matter (in the energy budget) during the collapse phase of megaparsec-scale structures.
\end{abstract}


%


\section{Introduction} \label{sec::intro}

The current paradigm of cosmology, the standard model or the $\Lambda$CDM model (Cold Dark Matter with a cosmological constant $\Lambda$), prescribes a globally homogeneous model universe composed of 68.3\% of dark energy, in the simplest and successful\footnote[1]{A disclaimer is in order here by referring to a number of existing observational `tensions', see \cite{tensions}.} case described by the cosmological constant $\Lambda$, 26.4\% of dark matter and 4.9\% of baryonic matter \cite{Planck2018VI}. As part of this model, the study of the formation of the large-scale structure is detached from the global expansion in the sense that the background model universe influences the growth of structure, but backreaction, i.e. the influence of structure growth on the universe model is suppressed by construction. In the $3+1$ formalism of general relativity, the averaged equations established in \cite{Buchert2000,Buchert2001,Buchert2008} describe the interaction of inhomogeneities with the averaged expansion through background-free volume averaging of the Einstein equations without approximations, and this includes cosmological backreaction.

For now, much effort has been put into the description of the dark energy through inhomogeneities, but very few papers have explored the link between dark matter and backreaction, e.g. \cite{Wiegand2010}. This is due to the large body of independent evidence for dark matter, from rotation curves of galaxies to the fluctuations in the Cosmic Microwave Background (CMB). Addressing all those issues with the same phenomenon remains difficult. This is discussed in section~\ref{sec::results_power_spectrum}.
This paper aims at quantifying the amount of `cosmological dark matter' which can be related to the geometrical inhomogeneities in the Universe. By the term \textit{cosmological dark matter} we refer to the phenomenon of dark matter limited to its effects on the formation and collapse of megaparsec-scale structures, especially its importance in the energy budget of these structures. We use the formalism of the averaged equations and a correspondence of the backreaction variables with an effective scalar field, called \textit{the morphon field} (related to morphological properties of structure \cite{Buchert2008}). This allows us to compare fundamental models of Scalar Field Dark Matter (henceforth SFDM), used in standard cosmology to describe clusters of galaxies and cosmological dark matter, with the effective description of the morphon field. By deriving the dynamical properties of collapsing domains of the Universe, this formalism also allows us to probe the effect of averaged curvature and kinematical backreaction on the process of collapse for generic domains.

The averaging formalism is presented in section \ref{sec::Back_morph} along with the morphon field. In section~\ref{sec::morph_SFDM}, the main principles behind SFDM models are presented. Sections~\ref{sec::Morph_SS} and \ref{sec::RZA} present a detailed analysis of the morphon and the dynamical properties of two inhomogeneous models: the exact class of scaling solutions in section~\ref{sec::Morph_SS}, and an approximate solution of Einstein's equations in the framework of a relativistic Lagrangian perturbation approach in section~\ref{sec::RZA}. In section~\ref{sec::results} the results are presented, firstly by comparing morphon fields with scalar fields from SFDM models, and secondly by analyzing the effect of backreaction on properties of collapsing domains.


\section{The Backreaction Context}
\label{sec::Back_morph}
\subsection{Averaged Einstein Equations} \label{sec::av_eq}

Spatially averaged Einstein equations have been introduced to generalize the Friedmann equations for homogeneous-isotropic distributions by taking into account the (matter and geometric) inhomogeneities of the Universe. The formalism used to derive them is the $3+1$ formalism in general relativity \cite{Gourgoulhon2012} that considers a slicing of the space-time manifold $\mathcal{M}$ into spacelike hypersurfaces $\Sigma_t$ of constant proper time $t$. Then, the spatial average of a space-time function $F$ is realized for each time $t$ on the hypersurface $\Sigma_t$ with the following operator:\footnote{The signature of the space-time metric is $\left(-1,1,1,1\right)$; $i,j,k$ run in $\left\{1,2,3\right\}$. We set $c\equiv 1$.}
\begin{eqnarray}
	\average{F\left(t, X^i\right)}{\CD} = \frac{1}{V_{\CD}\left(t\right)} \int_{\CD} F\left(t, X^i\right) \sqrt{\mathrm{det}\left(\gamma_{ij}\right)}\rd X^i \, ,
\end{eqnarray}
where $\CD$ is a comoving (Lagrangian) domain in the hypersurface on which the average is realized, $V_{\CD}\left(t\right) = \int_{\CD} \sqrt{\mathrm{det}\left(\gamma_{ij}\right)}\rd {X}^i$ is the volume of the domain $\CD$, assumed to be compact, and $\gamma_{ij}$ is the spatial metric of the hypersurface. Then, spatially averaging the Einstein equations and supposing that we are dealing with a dust (pressure-less) and irrotational fluid leads to the following set of cosmological equations:\footnote{We recall that these equations are covariantly defined, while they are written in proper time foliation (see \cite{foliations} for arguments in favour of this foliation choice, and \cite{astapierre} for the issue of covariance).}
\begin{subequations}
	\label{eq::averaged}
	\begin{eqnarray}
		\label{eq::Hamilton_constraint} H_{\CD}^2 &:=& \left(\frac{\dot{a}_\CD}{a_\CD}\right)^2 =  \frac{\kappa}{3}\average{\varrho}{\CD} - \frac{\average{\cal R}{\CD} + {\CQ_\CD}}{6} + \frac{\Lambda}{3} \, ; \\
		\label{eq::Ray_eq} \frac{\ddot{a}_{\CD}}{a_{\CD}} \, &=& -\frac{\kappa}{6} \average{\varrho}{\CD} + \frac{{\CQ_\CD}}{3} + \frac{\Lambda}{3} \, ,
	\end{eqnarray}
\end{subequations}
where $\kappa := 8\pi G$, the overdot denotes the covariant time-derivative, $\average{\varrho}{\CD} = {\varrho_{\initial\CD}/a_{\CD}^3}$ (with $\varrho_{\initial\CD}$ being a constant and where the subscript $\rm i$ stands for `initial'), is the average rest mass density of matter, $\average{\cal R}{\CD}$ is the average scalar curvature of the 3D--hypersurface in the domain $\CD$, and $\CQ_\CD$ is called the \textit{kinematical backreaction},
\begin{eqnarray}
	\CQ_\CD = \frac{2}{3}\average{\left(\theta - \average{\theta}{\CD}\right)^2}{\CD} - 2\average{\sigma^2}{\CD} \, , \label{eq::QD}
\end{eqnarray}
where $\theta$ is the trace of the expansion tensor $\Theta_{ij}$, and $\sigma^2 = \frac{1}{2} \sigma_{ij}\sigma^{ij}$ the squared rate of shear with $\sigma_{ij}$ the trace-free part of $\Theta_{ij}$; recall that vorticity is zero due to the foliation assumption. The domain-dependent function $a_{\CD}$ is the volume scale factor of the domain $\CD$, defined as $a_{\CD} = \left(V_{\CD}(t) / V_{\initial{\CD}}\right)^{1/3}$.

Equation~\eref{eq::Hamilton_constraint} is derived from averaging the energy constraint, and equation~\eref{eq::Ray_eq} from averaging the Raychaudhuri equation. Equations~\eref{eq::averaged} feature terms similar to the Friedmann equations for a homogeneous-isotropic model universe, but with additional backreaction terms: the kinematical backreaction $\CQ_\CD$ encodes non-local kinematical variance terms, and the averaged curvature $\average{\cal R}{\CD}$ that does not necessarily behave according to  the scale factor dependence expected in the standard model, i.e. $\propto a_{\cal D}^{-2}$. In order to fully compare the system of equations~\eref{eq::averaged} with the Friedmann equations, we isolate the backreaction variables and introduce the curvature deviation $\CW_\CD := \average{\CR}{\CD} - 6k_\initial{\CD} a_\CD^{-2}$, which probes the deviation of the average scalar curvature from the constant curvature model, $6k_\initial{\CD} a_\CD^{-2}$. Equation~\eref{eq::Hamilton_constraint} can then be written in the form:
\begin{equation}
	\label{eq::Hamilton_W} H_{\CD}^2 = \frac{\kappa}{3}\average{\varrho}{\CD} -\frac{k_\initial{\CD}}{a_\CD^2} - \frac{\CW_\CD + {\CQ_\CD}}{6}+ \frac{\Lambda}{3} \, .
\end{equation}
Equation~\eref{eq::Ray_eq} and the derivative of equation~\eref{eq::Hamilton_W}, together with the conservation law
\begin{equation}
\label{eq::conservation}
\average{\varrho}{\CD}^{\fdot} + 3 H_{\CD}\average{\varrho}{\CD} = 0 \, ,
\end{equation}
imply the integrability condition:
\begin{equation}
a_{\CD}^{-2}( a_{\CD}^{2}{\CW}_\CD )^{\fdot}\;+\;a_{\CD}^{-6}( a_{\CD}^{6}{\CQ}_{\CD} )^{\fdot} = 0 \, .
	\label{eq::integrability}
\end{equation}
In the set of equations $\lbrace$\eref{eq::Ray_eq}, \eref{eq::Hamilton_W}, \eref{eq::conservation}, \eref{eq::integrability}$\rbrace$ one equation is redundant.

In some papers (e.g. \cite{Wiegand2010, Roy2010, zimdahl:viscosity}) the $\CW_\CD$ and $\CQ_\CD$ terms are considered to be leading to an effect similar to dark energy. For instance, taking a positive backreaction $\CQ_\CD$ leads to volume acceleration (positive second derivative of the scale factor $a_\CD$) when the matter contribution in \Eref{eq::Ray_eq} becomes negligible against $\CQ_\CD$, thus mimicking a dark energy behaviour at the onset of large-scale structure formation. However, few studies have focussed on a possible interpretation of the curvature deviation and the backreaction as effective dark matter on cosmological scales. This article paves the way to such a description, using a correspondence between the backreaction variables and a scalar field, called the \textit{morphon field}, which can be linked to fundamental scalar field dark matter models. The morphon field is presented in the next subsection.

\subsection{The Morphon Field} \label{sec::morphon_field}

In view of the similarities between the general cosmological equations and the Friedmann equations, the curvature deviation and backreaction terms can be interpreted as being sourced by an effective cosmological fluid, called backreaction fluid, leading to a total effective energy density $\varrho_\mathrm{eff}$ and effective pressure $p_\mathrm{eff}$ \cite{Buchert2001,Buchert2008}:
\begin{subequations}
	\label{eq::fluid_eff}
	\begin{eqnarray}
		\varrho_\mathrm{eff} &:=& \average{\varrho}{\CD} -\frac{\CQ_\CD + \CW_\CD}{2\kappa} = \average{\varrho}{\CD} + \varepsilon_\mathrm{BF} \, ; \\
		p_\mathrm{eff} &:=& -\frac{\CQ_\CD - \frac{1}{3}\CW_\CD}{2\kappa} = p_\mathrm{BF} \, , 
	\end{eqnarray}
\end{subequations}
where $\varepsilon_\mathrm{BF}$ and $p_\mathrm{BF}$ are respectively the effective energy density and pressure of the backreaction fluid (BF). Equations~\eref{eq::Ray_eq}, \eref{eq::conservation} and \eref{eq::integrability} then assume Friedmannian form:
\begin{subequations}
	\label{eq::Buchert_eff}
	\begin{eqnarray}
		\frac{\ddot{a}_{\cal D}}{a_{\cal D}} \, &=& -\frac{\kappa}{6} \left(\varepsilon_\mathrm{eff}+ 3p_\mathrm{eff}\right) + \frac{\Lambda}{3} \, ; \\
		\dot{\varepsilon}_\mathrm{eff} &=& -3H_\CD\left(\varepsilon_\mathrm{eff} + p_\mathrm{eff}\right) \, , \label{eq::rho_eq_eff}
	\end{eqnarray}
with the energy constraint
\begin{eqnarray}
	\label{eq::Hamilton_eff}
	H_{\cal D}^2 = \frac{\kappa}{3}\varepsilon_\mathrm{eff} + \frac{k_\initial{\CD}}{a_\CD^2} + \frac{\Lambda}{3} \, .
\end{eqnarray}
\end{subequations}
As long as no assumption is made on the backreaction fluid, the system of equations~\eref{eq::Buchert_eff} is not closed. We then need to provide a dynamical equation of state that links the effective energy and pressure sources, possibly containing an explicit volume dependence through the volume scale factor.

Looking at the effective sources \eref{eq::fluid_eff}, we see that that the kinematical backreaction obeys a \textit{stiff equation of state}, i.e. $p_\mathrm{BF} = \varepsilon_\mathrm{BF}$ for $\CW_\CD = 0$, while the curvature deviation term obeys a \textit{curvature equation of state}, i.e. $p_\mathrm{BF} = -\frac{1}{3} \varepsilon_\mathrm{BF}$ for $\CQ_\CD = 0$ (with a sign change in front of $\CW_\CD = 0$), suggesting a correspondence with a scalar field \cite{Buchert2001,Buchert2006}: the backreaction fluid, with effective energy-momentum tensor $T_{\mu\nu}^\mathrm{BF} = \left(\varepsilon_\mathrm{BF} + p_\mathrm{BF}\right) \delta_\mu^0\delta_\nu^0 + p_\mathrm{BF} \, g_{\mu\nu}$, is expressed in this correspondence through an effective scalar field $\psi_\CD$ minimally coupled to a potential $U_\CD$,\footnote{See \cite{Madsen1987} for a maximally coupled scalar field.} with energy-momentum tensor $T_{\mu\nu}^{\psi_\CD} = \epsilon \dot{\psi}_\CD^2 \delta_{\mu\nu} + \left[\epsilon \frac{1}{2}\dot{\psi}_\CD^2 - U_\CD\left(\psi_\CD\right)\right] \, g_{\mu\nu}$. We obtain the following representation of the backreaction fluid density $\varrho_\mathrm{BF}$ and pressure $p_\mathrm{BF}$:
\begin{subequations}
	\label{eq::fluid_morph}
	\begin{eqnarray}
		\varepsilon_\mathrm{BF} = \epsilon \frac{1}{2}\dot{\psi}_\CD^2 + U_\CD\left(\psi_\CD\right) \, ; \\
		p_\mathrm{BF} = \epsilon \frac{1}{2}\dot{\psi}_\CD^2 - U_\CD\left(\psi_\CD\right) \, ,
	\end{eqnarray}
\end{subequations}
where $\epsilon = +1$ for a standard scalar field and $\epsilon = -1$ for a phantom field (with a negative kinetic energy density). Then we can interpret the backreaction fluid as a fluid model of an effective scalar field evolving in a potential, which leads to the correspondence:
\begin{subequations}
\label{eq::U_psi_dot}
\begin{eqnarray}
	\dot{\psi}^2_\mathcal{D} &=& -\epsilon\frac{\CW_\CD + 3{\cal Q_D}}{3\kappa} \, ; \label{eq::psi_dot} \\
	U_\mathcal{D} &=& \frac{-\CW_\CD}{3\kappa} \, . \label{eq::U}
\end{eqnarray}
\end{subequations}
The scalar field $\psi_\CD$ is called \textit{the morphon field}. The change of variables $(\CW_\CD, \CQ_\CD) \rightarrow (\psi_{\CD}, U_\CD)$ is not necessarily a diffeomorphism; there are cases where we have to change the morphon from a real to a phantom field, respectively from a phantom to a real field, each time $\CW_\CD + 3\CQ_\CD$ changes its sign from negative to positive, respectively from positive to negative. This issue will be discussed later in section~\ref{sec::RZA}.

In the framework of the morphon field, the integrability condition \eref{eq::integrability}
becomes the Klein-Gordon equation for the scalar field $\psi_\CD$:
\begin{eqnarray}
		 \ddot{\psi}_\CD &=& - 3H\dot{\psi}_\CD - \epsilon \frac{\rd U_\CD (\psi_\CD)}{\rd \psi_\CD} \, . \label{eq::KG_D}
\end{eqnarray}
The system of equations~\eref{eq::Buchert_eff} written in terms of the morphon field is formally equivalent to the Friedmann equations describing a model universe sourced by dust matter and a fundamental scalar field $\psi$ minimally coupled to a potential $U\left(\psi\right)$, e.g. \cite{Sahni2000}: 
\begin{subequations}
	\label{eq::Fried_psi}
	\begin{eqnarray}
		\frac{\ddot{a}}{a} \ \, \, &=& -\frac{\kappa}{3} \left(\frac{\varrho_{\rm m}}{2} + \epsilon\dot{\psi}^2 - U \right) \, ; \\
		 \ddot{\psi} &=& - 3H\dot{\psi} - \epsilon\frac{\rd U}{\rd \psi} \, ,
	\end{eqnarray}
with the energy constraint
\begin{eqnarray}
 		H^2 = \frac{\kappa}{3}\left(\varrho_{\rm m} + \epsilon \frac{1}{2}\dot{\psi}^2 + U \right) \, ,
\end{eqnarray}
\end{subequations}
where $\varrho_{\rm m}$ is the matter density of the homogeneous model universe.

As said before, the system of equations~\eref{eq::Buchert_eff} needs a closure condition to be solved. There exist different ways to define such a condition. The potential $U_\CD\left(\psi_\CD\right)$ of the morphon can be set. This is similar to what SFDM models do (see section~\ref{sec::morph_SFDM}) but with an effective approach in our case. In section~\ref{sec::Morph_SS}, closure is achieved with an ansatz on $\average{\CR}{\CD}\left(a_\CD\right)$ and $\CQ_\CD\left(a_\CD\right)$ called the \textit{the scaling solution}. In section~\ref{sec::RZA} closure is achieved by deriving $\CW_\CD$ and $\CQ_\CD$ from an explicit model for structure formation.

\subsection{Comparing SFDM and the morphon field}
\label{sec::mimick_DM}

Two approaches can be made to compare scalar fields from SFDM models and morphon fields from inhomogeneous models.

The first approach is to constrain the morphon with the fundamental models of SFDM. The potential of the morphon is chosen to be the same as for those models. In this case the system of equations describing the evolution of the scale factor $a_\CD$, equations~\eref{eq::Buchert_eff}, is the same as that in SFDM models (equations~\eref{eq::Fried_psi}). Then, setting the same initial conditions would lead to the same results in terms of the scale factor evolution. The difference resides in the interpretation of the scalar field: in the SFDM picture, it is fundamental and is linked to the expectation of existence of fundamental particles of dark matter; in the morphon picture, it is effective and is the result of backreaction. In this latter picture we can obtain an empirical estimate of the curvature needed to mimic \textit{cosmological dark matter}, as defined in the introduction. However, we do not have a physical justification, other than fitting the SFDM model, for the shape of the effective potential governing the inhomogeneities. This approach is described in the next section.

In the second approach, we infer the morphon field from an analytic solution of $\CW_\CD\left(t\right)$ and $\CQ_\CD\left(t\right)$ that describes structure formation. This approach allows us to give physical justification to the form of the potential, which can be directly compared to fundamental potentials employed in SFDM. This approach is firstly described in section~\ref{sec::Morph_SS} from a class of exact solutions of the general cosmological equations: the scaling solution. However, this analytic solution remains an ansatz used to close the system of equations~\eref{eq::Buchert_eff}. 
It will, however, give a first insight into the kind of dark matter potentials we expect from backreaction. In section~\ref{sec::RZA} we employ relativistic Lagrangian perturbation theory, allowing us to derive a morphon from a structure formation scenario, argued to be reasonably generic in \ref{appA}.

\section{Putting into Perspective Scalar Field Dark Matter models}
\label{sec::morph_SFDM}

In this section we shall briefly review SFDM models. Then, we will present the characteristics of the curvature resulting from the correspondence to the description of inhomogeneities by a morphonic SFDM model.

\subsection{Scalar Field Dark Matter models}
\label{sec::SFDM_th}

SFDM models are models of fundamental dark matter where the particles of dark matter can be modeled by a scalar field $\psi$ coupled to a potential $U\left(\psi\right)$. These models have been introduced to account for many observable issues related to dark matter, such as rotation curves of galaxies, structure formation, acoustic peaks of the CMB, and the evolution of cosmological parameters. In these models, dark matter can be considered either in a static description on scales of kpc to Mpc---because of their low mass, $\sim 10^{-22}$ eV, the particles of SFDM form Bose-Einstein condensates with sizes of some kpc, e.g. \cite{Matos2009a}---or, SFDM can be included in the dynamical description of cosmological dark matter in the Friedmann equations. It is this latter case we shall consider for our purpose.

Different forms for the potential $U\left(\psi\right)$ can be found in the literature. The main constraint on the potential is that it needs a minimum in order to be able to define a classical mass for the dark matter particles. The simplest form is the quadratic potential $U\left(\psi\right) = \frac{m_\mathrm{DM}^2}{2 \, \hbar^2} \, \psi^2$, with the dimension of $\psi$ being in $1/\sqrt{\kappa}$. A study of the resulting dynamical system of equations is made in \cite{Matos2009a, Urena2009}. However, \cite{Matos2009a} uses an ansatz for the Hubble parameter (Equation~(50) in \cite{Matos2009a}) which is not needed.\footnote{Equation~(50) in \cite{Matos2009a} assumes the Hubble parameter to be a power law of time: $H \propto t^{-n}$, with $n$ a free parameter. However, for purely dust-filled models, $n$ is not free and should be $1$. Or, just counting the needed number of equations, this assumption is redundant since the system is already closed \cite{Godart2016}.}
For a field with self-interaction, one can add a quartic term, with $U\left(\psi\right) = \frac{m_\mathrm{DM}^2}{2 \, \hbar^2} \, \psi^2 + \frac{\lambda}{4} \, \psi^4$, as in \cite{Matos2001}, or temperature-dependent terms as in \cite{Bernal2017}.
In \cite{Sahni2000, Matos2001, Matos2009a}, the authors assume a potential of the form $U\left(\psi\right) = U_0 \left[\cosh\left(\lambda\sqrt{\kappa} \, \psi \right) - 1\right]$ with $U_0 >0$, having an exponential behaviour for high values of $\psi$. Some motivations for this kind of potential can be found in particle physics and in inflationary scenarios (see \cite{Ferreira1998} for a study of a purely exponential potential).
In \cite{Matos2009b} the authors modify this $\cosh$-potential by adding the cosmological constant to its definition, leading to $U\left(\psi\right) = U_0 \left[\cosh\left(\lambda\sqrt{\kappa} \, \psi \right) - 1\right] + \Lambda/\kappa$ with $U_0>0$. This allows for the description of both dark matter and dark energy with the same potential.

The mass $m_\mathrm{DM}$ of the SFDM particles can be constrained by their density power spectrum, which presents a natural cut-off \cite{Hu2000, Matos2001}. The various studies on the topic (see \cite{Magana2012} for a brief review of SFDM models) agree on values of the mass $m_\mathrm{DM} \sim 10^{-22}\,$eV.

\subsection{Curvature from an effective SFDM model}
\label{sec::SFDM_Curvature}

As presented in section~\ref{sec::mimick_DM}, we can use this class of dark matter models to constrain the morphon field and the resulting backreaction terms. In this view the SFDM model is an effective model for the backreaction fluid. We here consider potentials describing only dark matter, without dark energy. The system of average equations~\eref{eq::Buchert_eff} is closed by fixing the potential. All the SFDM potentials presented in section~\ref{sec::SFDM_th} are positive potentials. Then, from equation~\eref{eq::U}, this implies a negative curvature deviation $\CW_\CD$ for all times, with a zero curvature at the extremum of the potential. As we shall learn in the next sections, from analytical scaling solutions (section~\ref{sec::Morph_SS}) and from an approximate solution for structure formation (section~\ref{sec::RZA}), we expect the average curvature to be positive with a phantom scalar field in regions where dark matter dominates. We will highlight these differences between the fundamental models and the analytical solutions for inhomogeneities in section~\ref{sec::results_SFDM}.

\section{Morphon Field from a Scaling Solution}
\label{sec::Morph_SS}

This section presents an example for a class of exact solutions of the average equations for which we can explicitly derive the potential of the morphon field.

\subsection{Classes of Scaling Solutions}
\label{sec::SS}

In this approach we express $\CW_\CD$ and $\CQ_\CD$ in terms of power laws of the volume scale factor $a_{\cal D}$, $\CW_\CD = \CW_\initial{\CD} a_\CD^n$ and $\CQ_\CD = \CQ_\initial{\CD} a_\CD^p$. The integrability condition~\eref{eq::integrability} distinguishes two classes of solutions \cite{Buchert2006,Roy:instability}:
\begin{itemize}
	\item[(i)] $\left(n,p\right) = \left(-2,-6\right)$. In this case, the curvature follows the evolution law $\propto a_\CD^{-2}$ as in the Friedmann equations. The backreaction term being $\propto a_\CD^{-6}$; it becomes rapidly negligible compared with $\CW_\CD$. This degenerate case mirrors the situation in the standard model: it decouples structure formation and the background model universe.
	\item[(ii)] $n=p$. In this case, supposing $n\not=-6$,\footnote{In the case $n=p=-6$ we have $\CW_\initial{\CD} = 0$ and $r$ is not defined.} we can rewrite $\CQ_\CD$ as $\CQ_\CD = r \CW_\initial{\CD} a_\CD^n$, with $r := 
	\CQ_\initial{\CD}/ \CW_\initial{\CD} = -\frac{n+2}{n+6}$ (or~$n = -2\frac{1+3r}{1+r}$).
\end{itemize}
In the following calculations we shall consider the second generic case 
and derive the effective potentials of the scaling solutions. For $n\not= -3$ and $n\not=0$,\footnote{The case $n=0$ is equivalent to a scale-dependent cosmological constant; we have the relations $\CW_\CD = 3\Lambda_\CD$ and $\CQ_\CD = \Lambda_\CD$.} two cases appear depending on the sign of $\gamma^{\initial{\CD}}_{\CW\CQ\mathrm{m}} := \left(\Omega^{\initial{\CD}}_\CW + \Omega^{\initial{\CD}}_\CQ\right) / \Omega^\initial{\CD}_\mathrm{m}$ (the following equations are corrected versions of \cite{Buchert2006}):\footnote{\textit{Erratum}: In \cite{Buchert2006}, $\psi^{+}_\CD$ is wrong by a factor of $2$, and $U_\mathcal{D}$ by a factor of $(1+r)^2$.}
\begin{itemize}
	\item[(i)] If $\gamma^{\initial{\CD}}_{\CW\CQ\mathrm{m}} < 0$:
\end{itemize}
\begin{subequations}
	\label{eq::phi_U_sin_bis}
	\begin{eqnarray}
		\psi_\mathcal{D}\left(a_\mathcal{D}\right) = \psi_\CD^{+} \, \arcsin\left(a_\mathcal{D}^\frac{n+3}{2} \sqrt{-\gamma^{\initial{\CD}}_{\CW\CQ\mathrm{m}}}\right) + \psi_\mathrm{c} \, ; \label{eq::SS_gamma<0_psi} \\
		U_\mathcal{D}\left(\psi_\mathcal{D}\right) = -\frac{H^2_{\initial{\CD}}\left(n+6\right)}{2\kappa\left(1+\gamma^{\initial{\CD}}_{\CW\CQ\mathrm{m}}\right)}\left(-\gamma^{\initial{\CD}}_{\CW\CQ\mathrm{m}}\right)^{\frac{3}{n+3}} \, \sin^{\frac{2n}{n+3}}\left(\frac{\psi_\CD - \psi_\mathrm{c}}{\psi_\CD^{+}}\right) \, , \label{eq::SS_gamma<0_U}
	\end{eqnarray}
\end{subequations}
where $\psi_\mathrm{c}$ is an integration constant and
\begin{eqnarray}
	\nonumber \psi_\CD^{+} = \pm \sqrt{\frac{\epsilon n}{\kappa}}\frac{2}{n+3} \, ,
\end{eqnarray}
along with the condition $\epsilon = +1$, if $n > 0$ (real field) and $\epsilon = -1$, if $n < 0$ (phantom field). Equation~\eref{eq::SS_gamma<0_psi} is well-defined initially because $-\gamma^{\initial{\CD}}_{\CW\CQ\mathrm{m}} < 1$ due to $\Omega^\initial{\CD}_\mathrm{m} >0$. Moreover, if $n+3>0$, $a_\CD$ has an upper limit $a_{\CD_{\rm ta}}$, corresponding to a turn-around (hence the subscript `$\rm ta$') of the domain scale factor with $a_{\CD_{\rm ta}} := (-\gamma^{\initial{\CD}}_{\CW\CQ\mathrm{m}})^{-1 / (n+3)}$, and if $n+3<0$, this limit is a lower limit. 
\begin{itemize}
	\item[(ii)] If $\gamma^{\initial{\CD}}_{\CW\CQ\mathrm{m}} > 0$:
\end{itemize}
\begin{subequations}
	\label{eq::phi_U_sh_bis}
	\begin{eqnarray}
		\psi_\CD\left(a_\CD\right) = \psi_\CD^{-} \, \mathrm{arsinh}\left(a_\CD^\frac{n+3}{2} \sqrt{\gamma^{\initial{\CD}}_{\CW\CQ\mathrm{m}}}\right) + \psi_\mathrm{c} \, ; \label{eq::SS_gamma>0_psi} \\
		U_\CD\left(\psi_\CD\right) = \frac{H^2_{\initial{\CD}}\left(n+6\right)}{2\kappa\left(1+\gamma^{\initial{\CD}}_{\CW\CQ\mathrm{m}}\right)}\left(\gamma^{\initial{\CD}}_{\CW\CQ\mathrm{m}}\right)^{\frac{3}{n+3}} \, \sinh^{\frac{2n}{n+3}}\left(\frac{\psi_\CD - \psi_\mathrm{c}}{\psi_\CD^{-}}\right) \, , \label{eq::SS_gamma>0_U}
	\end{eqnarray}
\end{subequations}
with
\begin{eqnarray}
	\nonumber \psi_\CD^{-} = \pm \sqrt{\frac{-\epsilon n}{\kappa}}\frac{2}{n+3} \, ,
\end{eqnarray}
along with the condition $\epsilon = +1$, if $n < 0$, and $\epsilon = -1$, if $n > 0$.

In the case where $n=-3$ the solution reads:
\begin{subequations}
	\label{eq::phi_U_n=-3}
	\begin{eqnarray}
		\hspace{-2.5cm} \psi_\CD\left(a_\CD\right) = \mp \sqrt{\frac{\epsilon \, \gamma^{\initial{\CD}}_{\CW\CQ\mathrm{m}}}{3\kappa\left(1+\gamma^{\initial{\CD}}_{\CW\CQ\mathrm{m}}\right)}} \log\left(a_{\cal D}^{-3}\right) + \psi_\mathrm{c} \, ; \\
		\hspace{-2.5cm} U_\CD\left(\psi_\CD\right) = \frac{H^2_{\initial{\CD}}\left(n+6\right)}{2\kappa \left(1+\gamma^{\initial{\CD}}_{\CW\CQ\mathrm{m}}\right)} \gamma^{\initial{\CD}}_{\CW\CQ\mathrm{m}} \exp{\left[\ \mp \left(\psi_\CD - \psi_\mathrm{c}\right)\sqrt{\frac{3 \kappa\left(1+\gamma^{\initial{\CD}}_{\CW\CQ\mathrm{m}}\right)}{\epsilon \, \gamma^{\initial{\CD}}_{\CW\CQ\mathrm{m}}}}\ \right]} \, , \label{eq::SS_n=-3}
	\end{eqnarray}
\end{subequations}
with the condition $\epsilon = +1$, if $\gamma^{\initial{\CD}}_{\CW\CQ\mathrm{m}} > 0$, and $\epsilon = -1$, if $\gamma^{\initial{\CD}}_{\CW\CQ\mathrm{m}} < 0$.

The form of the potentials \eref{eq::SS_gamma<0_U}, \eref{eq::SS_gamma>0_U} and \eref{eq::SS_n=-3} is a consequence of the power law ansatz.  But, the inverse statement is incorrect: setting the potential of the morphon to be one of the possibilities  \eref{eq::SS_gamma<0_U}, \eref{eq::SS_gamma>0_U} or \eref{eq::SS_n=-3} does not necessarily lead to a power law behaviour of $\CW_\CD$ and $\CQ_\CD$. For instance, the potential \eref{eq::SS_gamma<0_U} can lead to oscillations of the scalar field around its maximum, which is not the case with the scaling solution.
The cases $\gamma^{\initial{\CD}}_{\CW\CQ\mathrm{m}} > 0$ and $n=-3$ correspond to an ever expanding domain. The case $\gamma^{\initial{\CD}}_{\CW\CQ\mathrm{m}} < 0$ corresponds to a collapsing domain. This collapse is time-symmetric with a turn-around arising at $a_\CD = a_{\CD_{\rm ta}}$, and a collapse at  $a_\CD = 2 \, a_{\CD_{\rm ta}}$.

\subsection{Scaling Solutions and Dark Matter}
\label{sec::SFDM_SS}

We can now compare the potentials derived from scaling solutions to potentials of SFDM models. In particular, we will compare the effective mass that we can derive from the scaling potentials. Even if these potentials are not exactly the same as those used in SFDM models, the mass will give a characteristic evolution time around its maximum which can then be compared with the fundamental models.

As explained in section~\ref{sec::SFDM_th}, if we want to mimic dark matter with a morphon field in the same way as in SFDM models, we need a quadratic dominant term in the Taylor expansion of the potential, i.e. $U_{\cal D}\left(\psi\right) = U_0 + \frac{m_{\rm eff}^2}{2 \, \hbar^2}\psi_\CD^2 + \mathcal{O}\left(\psi_\CD^3\right)$. This second-order term represents the effective mass of the morphon.

In the case $\gamma^{\initial{\CD}}_{\CW\CQ\mathrm{m}} > 0$, i.e. $\Omega_\mathrm{m}^\initial{\CD} < 1$, the potential \eref{eq::SS_gamma>0_U} has no quadratic minimum, unless if we assume the scaling exponent $n$ to be infinite. Thus, no effective mass can be attributed to this morphon.

The case $\gamma^{\initial{\CD}}_{\CW\CQ\mathrm{m}} < 0$, i.e. $\Omega_\mathrm{m}^\initial{\CD} > 1$, has quadratic extrema. They correspond to maxima or minima depending on $n$. We consider $n<0$, because it is physically more relevant, since we expect the absolute value of the curvature to decrease (respectively increase) with an increase (respectively decrease) of the volume of the domain $\CD$. Then, according to section~\ref{sec::SS}, the morphon field is a phantom field.
In order to define an effective mass for this field, we have to consider the evolution equation for a general phantom field $\psi_\mathrm{phant}$, without expansion. This equation is the following:
\begin{eqnarray}
	\ddot{\psi}_\mathrm{phant} - \frac{\rd U}{\rd \psi_\mathrm{phant}} = 0 \, . \label{eq::KG_phant}
\end{eqnarray}
We infer from this equation that phantom fields are stable at the maxima of their potential \cite{Carroll2003}. With a potential of the form $U_\mathrm{phant}\left(\psi_\mathrm{phant}\right) = U_0 - \textstyle{\frac{m_\mathrm{phant}^2}{2 \, \hbar^2}}\psi_\mathrm{phant}^2$ we retrieve the classical Klein-Gordon equation for a particle of mass $m_\mathrm{phant}$. Then $m_\mathrm{phant}$ can be interpreted as the mass of the phantom field.

Thus, we consider the morphon field defined by the equations~\eref{eq::phi_U_sin_bis} and we search for maxima of its potential to be able to derive an effective mass $m_{\rm eff}$. In the domain of definition of $\psi_\CD$, given by equation~\eref{eq::SS_gamma<0_psi}, the potential \eref{eq::SS_gamma<0_U} has one extremum in $(\psi_\CD - \psi_\mathrm{c})/\psi_\CD^{+} = \frac{\pi}{2}$. It is a maximum only if $n \in ]-\infty,-6[\ \bigcup\ ]-3,0[$. For $n \in ]-\infty,-6[$ the average curvature deviation is negative and for $n \in ]-3,0[$ it is positive.

Hence, for $\gamma^{\initial{\CD}}_{\CW\CQ\mathrm{m}}<0$ and $n \in ]-\infty,-6[\ \bigcup\ ]-3,0[$, we can derive an effective mass $m_{\rm eff}$ of the morphon field resulting from the scaling solutions:
\begin{eqnarray}
	m_{\rm eff} = \hbar H_\initial{\CD} \sqrt{\frac{\left(n+3\right)\left(n+6\right)\left(-\gamma^{\initial{\CD}}_{\CW\CQ\mathrm{m}}\right)^{\frac{3}{n+3}}}{4\left(1+\gamma^{\initial{\CD}}_{\CW\CQ\mathrm{m}}\right)}} \, . \label{eq::SS_mass}
\end{eqnarray}
Realistic estimates for $m_{\rm eff}$ will be given in section~\ref{sec::results_SFDM}. The potential at this maximum can be written as:
\begin{eqnarray}
	U_\CD = -\frac{H^2_\initial{\CD} \left(n+6\right)\left(-\gamma^{\initial{\CD}}_{\CW\CQ\mathrm{m}}\right)^{\frac{3}{n+3}}}{2\kappa\left(1+\gamma^{\initial{\CD}}_{\CW\CQ\mathrm{m}}\right)} - \frac{m_{\rm eff}^2}{2 \, \hbar^2}\left(\Delta\psi_\CD\right)^2 + \mathcal{O}\left(\Delta\psi_\CD\right)^4 \, ,
\end{eqnarray}
where $\Delta\psi_\CD = \psi_\CD - \left(\frac{\pi}{2} \, \psi_\CD^+ + \psi_{\rm c}\right)$. It is important to note here that this result is only valid for $\gamma^{\initial{\CD}}_{\CW\CQ\mathrm{m}}<0$, i.e. $\Omega_\mathrm{m}^\initial{\CD} > 1$. Thus, the domain $\CD$ cannot represent the global evolution of the model universe. It can, however, represent an overdense subdomain with respect to the average rest mass density of the Universe. In this case we allow  for $\Omega_\mathrm{m}^\initial{\CD} > 1$. This is coherent with the expected behaviour of dark matter which mainly gathers in overdense regions of the Universe.  Then, as pointed out in \cite{Wiegand2010}, we expect the overdense regions to have a positive average scalar curvature. This would restrict the range of the scaling exponent to the interval $n \in ]-3,0[$. This range is in agreement with the partitioning approach made in \cite{Wiegand2010}, where the authors assume an extrapolation of the leading perturbative result for the backreaction variables, $n=-1$, for the two different unions of domains considered, overdense and underdense.

Although in this section we closed the system of equations~\eref{eq::Buchert_eff} without fixing the potential, we still have no physical justification for this closure, at least if we extrapolate to a time and a scale where scaling solutions, e.g. motivated by leading-order terms, do not capture the physical reality of the collapse. An approximate solution for structure formation using a relativistic Lagrangian perturbation approach will provide such a justification. We turn to this approach in the next section.

\section{Morphon Field from the Relativistic Lagrangian Perturbation Theory}
\label{sec::RZA}

In this section we derive the morphon field resulting from the backreaction and curvature deviation functionals of the Relativistic Zel'dovich Approximation (RZA) presented in \cite{RZA_1, RZA_2} (and references therein).

\subsection{The Relativistic Zel'dovich Approximation}
\label{sec::TheRZA}

In this approximation, $a_\CD$, $\CW_\CD$ and $\CQ_\CD$ are derived as nonlinear deviations from a fixed background model universe in the class of Friedmann--Lema\^\i tre solutions (Einstein--de Sitter model for instance, henceforth EdS for short).\footnote{For the corresponding background-independent equations, see \cite{Roy2012}.} Since this model is background-dependent, in contrast to our general setup, the amount of backreaction of the inhomogeneities on the global evolution of the model universe is suppressed on large scales. Hence, the predicted behaviour of $\CW_\CD$ and $\CQ_\CD$ from the RZA model provides a reliable probe only on relatively small and intermediate scales, roughly in between $5-50$Mpc. We therefore specify the domain $\CD$ to matter-dominated domains on this range of scales, and abbreviate it by $\CM$ hereafter. At those scales we expect the dark energy to play a negligible to subdominant role. Thus, if we derive the morphon field resulting from the RZA, this morphon will be assumed to probe effects of the inhomogeneities in this range of scales, i.e. we shall consider it as a pure dark matter morphon, and not a morphon mixing dark matter and dark energy. The latter case should be considered if we want to describe both dark components with backreaction effects and curvature deviation simultaneously within a single model universe. In the present framework, the physical phenomenon at the origin of the dark matter and dark energy is the same, thus justifying a single morphon field. We consider this unified treatment to lie beyond the scope of the present paper. A possible way of describing the general situation is provided by the general multi-scale approach of\cite{BuchertCarfora2008,Wiegand2010}; for special bi-scale modeling see also \cite{Rasanen2006,NazerWiltshire,Lavinto,Roukema}.

The basic principle behind this relativistic generalization of the Zel'dovich Approximation is to write the \textit{spatial} metric $g_{ij}$ it terms of \textit{spatial} co-frames:
\begin{eqnarray}
	g_{ij} = G_{ab} \, {\eta^a}_i{\eta^b}_j \, ,
\end{eqnarray}
where Gram's matrix $G_{ab}\left(\initial{t},X^k\right)$ is a constant of time and corresponds to the initial spatial metric, and the ${\eta^a}_i\left(t,X^k\right)$ are the spatial co-frames (triads),
with ${\eta^a}_i\left(\initial{t},X^k\right) = {\delta^a}_i$. Then, the background-dependent perturbation is realized on the co-frames with the perturbation matrix $P^a_i \left(t,X^k\right)$ as follows:
\begin{eqnarray}
	\eta^a_{\ i}\left(t,X^k\right) = a\left(t\right) \, \left[\delta^a_{\ i} + P^a_{\ i} \right] \, ,
\end{eqnarray}
where the background is imposed in the global scale factor $a(t)$. At first order in $P^a_{\ i}$ the co-frames are (see \cite{RZA_2} for more details):
\begin{eqnarray}
	\eta^a_{\ i} \left(t,X^k\right) = a\left(t\right) \, \left[\delta^a_{\ i} + \xi\left(t\right) \, {\dot P}^a_{\ i}\left(\initial{t},X^k\right)\right] \, , \label{eq::pert_co-frame}
\end{eqnarray}
where $\xi(t) = [q(t) - \initial{q}]/\initial{\dot{q}}$ with $q(t)$ being the \textit{leading growing mode} solution of the equation
\begin{eqnarray}
 \ddot{q}\left(t\right) + 2 \, \frac{\dot{a}\left(t\right)}{a\left(t\right)} \, \dot{q}\left(t\right) + \left(3 \, \frac{\ddot{a}\left(t\right)}{a\left(t\right)} - \Lambda\right)q\left(t\right) = 0 \, , \label{eq::RZA_q}
\end{eqnarray}
with $\Lambda$ the cosmological constant of the background.

Here, the extrapolation philosophy of the averaged Zel'dovich approximation \cite{BKS2000} is adopted in the relativistic case \cite{RZA_1,RZA_2}, specified to the domain of validity of RZA, $\CM$, where $a_\CM$, $\CW_\CM$ and $\CQ_\CM$ are calculated as functionals of the perturbed co-frames \eref{eq::pert_co-frame} without performing any other approximations. Then, different formulas can be obtained for each field variable, depending on the equation we use to derive it.\footnote{This degeneracy is a result of the fact that we deal with an extrapolation of a perturbative approximation whose quality can be tested by comparing the different formulas (see \cite{zeldovich,doroshkevich} for the Newtonian framework). We are not interested in a refined quantitative investigation of the Relativistic Zel'dovich Approximation in this paper, but the reader should keep in mind that we deal with approximate expressions. As an example, the volume scale factor $a_\CM$ can be derived either from the continuity equation (as adopted here), or through time-integration of the volume expansion or volume acceleration laws.}

We here choose to derive the approximate functional for the scale factor $a_\CM$ from the continuity equation, which allows to exactly conserve the sum of the rest masses of fluid elements in the domain $\CM$. Using the spatial average of equations~(41) and (42) in \cite{RZA_2}, we have:
\begin{equation}
	a_\CM = a(t) \left( 1+\xi (t) \laverage{\inI}{\CM} + \xi^2 (t) \laverage{\inII}{\CM} + \xi^3 (t) \laverage{\inIII}{\CM} \right)^{1/3} \, . \label{eq::RZA_aD}
\end{equation}
$\rm I$, $\rm II$ and $\rm III$ denote the three principal scalar invariants of the matrix ${\dot P}^a_{\ i}\ $.

The kinematical backreaction term $\CQ_\CM$, derived from its definition in terms of the expansion tensor \eref{eq::QD} (equation~(50) in \cite{RZA_2}) is the following:
\begin{eqnarray}
\label{Qfunctional}
	\nonumber \CQ_\CM = \frac{\dot{\xi}^2 \left(\gamma_1 + \xi\gamma_2+\xi^2\gamma_3\right)}{\left(1+\xi\laverage{\inI}{\CM} +\xi^2\laverage{\inII}{\CM} + \xi^3\laverage{\inIII}{\CM} \right)^2} \, ,
\end{eqnarray}
with
\begin{eqnarray}
	\label{eq::RZA_QD}
	\cases{
		\gamma_1 := 2\laverage{\inII}{\CM} - \frac{2}{3}\laverage{\inI}{\CM}^2 \, ; \\
		\gamma_2 := 6\laverage{\inIII}{\CM} - \frac{2}{3}\laverage{\inII}{\CM}\laverage{\inI}{\CM} \, ; \\
		\gamma_3 := 2\laverage{\inI}{\CM}\laverage{\inIII}{\CM} - \frac{2}{3}\laverage{\inII}{\CM}^2 \, .
		}
\end{eqnarray}
Equation~\eref{eq::RZA_QD} enjoys the property of covering subcases of exact solutions of the Einstein equations such as locally plane-symmetric solutions and a subclass of averaged spherically symmetric LTB (Lema\^\i tre-Tolman-Bondi) solutions (see \cite{RZA_2} and \cite{Buchert2011}).

For the average curvature deviation $\CW_\CM$, \cite{RZA_2} describes three ways of computation: (i) from the averaged energy constraint and the Raychaudhuri equation (equation~(54) in \cite{RZA_2}); (ii) from the integrability condition; (iii) from the geometric definition of the curvature in terms of derivatives of the spatial metric. All possibilities correspond to approximate estimates of averaged curvature, and they all agree in the limit where the Lagrangian first-order deformation reliably describes the Lagrangian linear regime. In the remainder of this article, the first choice is made, i.e. the approximate functional for $\CW_\CM$ is given by (for an evaluation of this choice see [Appendix B] in \cite{RZA_2}):
\begin{eqnarray}
	\nonumber \mathcal{\CW_\CM} = \frac{\dot{\xi}^2 \left(\tilde\gamma_1 + \xi\tilde\gamma_2+\xi^2\tilde\gamma_3\right)}{1+\xi\laverage{\inI}{\CM} +\xi^2\laverage{\inII}{\CM} + \xi^3\laverage{\inIII}{\CM}} + 6\left(\frac{k}{a^2} - \frac{k_\initial{\CM}}{a^2_{\CD}}\right) \, ,
	\label{Wfunctional}
\end{eqnarray}
with
\begin{eqnarray}
	\label{eq::RZA_WD}
	\cases{
		\tilde\gamma_1 := -2\laverage{\inII}{\CM} - 12\laverage{\inI}{\CM}\frac{H}{\dot{\xi}} - 4\laverage{\inI}{\CM}\frac{\ddot{\xi}}{\dot{\xi}^2} \, ; \\
		\tilde\gamma_2 := -6\laverage{\inIII}{\CM} - 24\laverage{\inII}{\CM}\frac{H}{\dot{\xi}} - 8\laverage{\inII}{\CM}\frac{\ddot{\xi}}{\dot{\xi}^2} \, ; \\
		\tilde\gamma_3 := -36\laverage{\inIII}{\CM}\frac{H}{\dot{\xi}} - 12\laverage{\inIII}{\CM}\frac{\ddot{\xi}}{\dot{\xi}^2} \, ,}
\end{eqnarray}
where $a$, $H$ and $k$ are, respectively, the scale factor, the expansion rate and the initial homogeneous curvature constant of the background model universe.\\

In the following subsections we will derive the morphon field from the RZA in a restricted subcase on the principal scalar invariants. In the weakly nonlinear regime, well before shell-crossing singularities develop, the first principal scalar invariant is dominating  \cite{BKS2000}. Looking at sufficiently large domains we can learn about important properties of the morphon potential in the 
collapsing phase by neglecting the \emph{initial} second and third principal scalar invariants, i.e. by taking $\laverage{\inI}{\CM} \not= 0$ and $\laverage{\inII}{\CM} = 0 = \laverage{\inIII}{\CM}$. In \ref{appA} we give arguments in support of this simplifying assumption. This subcase allows us to find analytic solutions for the morphon potential. It has also the advantage of exactly conserving the averaged energy constraint and the integrability condition, as explained in section~IV.B. of \cite{RZA_2}. This strengthens the results which will be presented in section~\ref{sec::results_Omegas}.

$\laverage{\inI}{\CM}$ is directly linked to the initial density perturbation with respect to the background \cite{BKS2000}: if $\laverage{\inI}{\CM}<0$, the domain $\CM$ is overdense with respect to the mean density imposed by the background; if $\laverage{\inI}{\CM}>0$, the domain $\CM$ is underdense (not considered here); if $\laverage{\inI}{\CM}=0$, the density of $\CM$ happens to be the same as that of the background.

In section~\ref{sec::RZA_EdS} we derive the morphon field for an EdS background. However, dark energy is not accounted for in this background model. Thus, we derive the morphon also for a $\Lambda$CDM background in section~\ref{sec::RZA_LCDM} to obtain a background evolution in conformity with the standard model with a dark energy component; recall that we concentrate our attention to the collapsing domains on intermediate scales $\CM$. It turns out that  the results for the morphon field are similar to the EdS case up until the collapse phase, showing the robustness of the scalar field language in our context. We use the solutions for the time-functions in \cite{Bildhauer1992} in both cases.

\subsection{RZA Morphon---EdS Background}
\label{sec::RZA_EdS}
In the case of an EdS background we have:
\begin{equation}
	a\left(t\right) = \left(\frac{t}{\initial{t}}\right)^{\frac{2}{3}} \quad ; \quad
	\xi\left(t\right) = \left[a\left(t\right) - 1\right] \frac{1}{\initial{H}} \, , \label{eq::RZA_EdS_xi}
\end{equation}
where $\initial{t} = 2/(3\initial{H})$, with $\initial{H}$ the initial Hubble parameter of the background. From equation~\eref{eq::RZA_aD} the volume scale factor $a_\CM$ is expressed as, assuming $\laverage{\inI}{\CM} \not= 0$, $\laverage{\inII}{\CM} = 0$ and $\laverage{\inIII}{\CM} = 0$:
\begin{eqnarray}
	a_\CM\left(a\right) = a \left(\frac{a-3\cstEdS}{1-3\cstEdS}\right)^{1/3} \, , \label{eq::RZA_100_aD}
\end{eqnarray}
where $\cstEdS :=  \frac{1}{3}(1 - \initial{H} / \laverage{\inI}{\CM})$ is a constant (E stands for `EdS') which can be seen as a (scale-dependent) critical scale factor. This notation was chosen because the critical values of the global scale factor $a$ will depend on $\cstEdS$. This leads to
\begin{eqnarray}
	H_\CM\left(a\right) = \frac{\initial{H}}{3 \, a^{3/2}}\frac{4a-9\cstEdS}{a-3\cstEdS} \, . 
	\label{eq::RZA_100_HD}
\end{eqnarray}
From the expression for $H_\CM$, we see that for $\cstEdS<1/3$ the domain $\CM$ is ever expanding (not considered here), and for $\cstEdS>1/3$, the domain $\CM$ is collapsing. The turn-around, i.e. the beginning of the collapse, arises at $a = a_{\rm ta} := \frac{9}{4} \, \cstEdS$. From equations~\eref{eq::RZA_QD}, \eref{eq::RZA_WD} and \eref{eq::RZA_EdS_xi} we have:
\begin{eqnarray}
	\CQ_\CM = \frac{-2\initial{H}^2}{3a\left(a - 3\cstEdS\right)^2} \quad ; \quad
	\CW_\CM = \frac{-10\initial{H}^2}{a^2\left(a-3\cstEdS\right)} \, . \label{eq::EdS_WQ_D}
\end{eqnarray}
We introduce the adimensional, reduced potential 
\begin{equation}
\uEdS := \frac{3\kappa}{6\initial{H}^2} U^{\mathrm{E}}_\CM = -\frac{\mathcal{\CW_\CM}}{6\initial{H}^2} \ .
\end{equation} 
Then, from equations~\eref{eq::U_psi_dot} we obtain:
\begin{equation}
	\uEdS = \frac{5/3}{a^2\left(a - 3\cstEdS\right)}\ \quad ; \quad 
	\partial_a \psiEdS = \pm\sqrt{\frac{2}{\kappa}}\sqrt{\frac{\epsilon\left(2a-5\cstEdS\right)}{a\left(a-3\cstEdS\right)^2}} \ ,
	\label{eq::da_psi_100}
\end{equation}
recalling that we take $\initial{a} = 1$.

Depending on the value of the critical scale factor $\cstEdS$, the integration of equation~\eref{eq::da_psi_100} leads to two different cases: ($C_1$) with $\cstEdS < 1/3$, i.e. $\laverage{\inI}{\CM}>0$, corresponding to an underdense domain $\CM$ with respect to the background, and ($C_2$) with $\cstEdS > 1/3$, i.e. $\laverage{\inI}{\CM}<0$, corresponding to an overdense domain $\CM$.

In the case ($C_1$) we have an ever expanding domain with a real field, $\epsilon = +1$. The scalar field is given by
\begin{eqnarray}
\fl \psiEdS = \pm \frac{2\sqrt{2}}{\sqrt{3\kappa}} \left[ - \mathrm{arcoth}\left(\sqrt{\frac{2a-5\cstEdS}{a/3}}\right) + \sqrt{6} \, \log\left(\sqrt{2a} + \sqrt{2a-5\cstEdS}\right) \right] + \psi_\mathrm{c} \, ,
\end{eqnarray}
where $\psi_\mathrm{c}$ is an integration constant. The inverse of the function $\psi_\CM\left(a\right)$, needed to compute the potential $\uEdS (\psiEdS )$, is not analytical. We can, however, infer its asymptotic behaviour: for $a \rightarrow +\infty$, $\uEdS \propto \exp(\mp\frac{3}{2}\sqrt{\kappa} \, \psiEdS)$. The global shape of this potential is not the one of a typical dark matter potential of SFDM.  It has no extremum, thus no effective mass can be associated to it. It diverges at a finite value of the scalar field and tends to zero for infinite values of this field. This last point is similar to the `early potential' of \cite{Ferreira1998}. This potential is depicted in figure~\ref{fig::morphon_RZA_LCDM} for $\cstEdS=-200$ and $\cstEdS=-300$. These values were chosen for reasons of visibility. We will not consider anymore this case as it does not correspond to a collapsing overdense domain.

In the case ($C_2$) we have a collapsing domain and the expression for $\psiEdS$ depends on the values of $a$ and $\cstEdS$. For $a < \frac{5}{2} \cstEdS$ and $\cstEdS > \frac{2}{5}$,\footnote{With $\frac{1}{3} < \cstEdS < \frac{2}{5}$, the solution is initially given by equation~\eref{eq::RZA_100_EdS_C2_2} because $\initial{a} = 1 > \frac{5}{2} \cstEdS$. This corresponds to a domain that is initially set to collapse, with $H_\initial{M} <0$.} i.e. $-5 < [\laverage{\inI}{\CM} / \initial{H}] < 0$, the morphon is a phantom field, $\epsilon = -1$, and we have:
\begin{eqnarray}
\fl \psiEdS = \pm \frac{2\sqrt{2}}{\sqrt{3\kappa}} \left[ - \mathrm{arccot}\left(\sqrt{\frac{-2a+5\cstEdS}{a/3}}\right) + \sqrt{6} \, \mathrm{arccot}\left(\sqrt{\frac{-2a + 5\cstEdS}{2a}}\right) \right] + \psi_\mathrm{c} \, ,
\end{eqnarray}
with an integration constant $\psi_\mathrm{c}$. For $5\cstEdS/2 < a < 3\cstEdS$, the morphon is a real field, $\epsilon = +1$, and we instead have:
\begin{eqnarray}
\fl \psiEdS = \pm \frac{2\sqrt{2}}{\sqrt{3\kappa}} \left[ \mathrm{artanh}\left(\sqrt{\frac{2a-5\cstEdS}{a/3}}\right) - \sqrt{6} \, \log\left(\sqrt{2a} + \sqrt{2a-5\cstEdS}\right) \right] + \psi_\mathrm{cc} \, , \label{eq::RZA_100_EdS_C2_2}
\end{eqnarray}
with another integration constant $\psi_\mathrm{cc}$. For $\psiEdS$ to be continuous at $a = \frac{5}{2} \cstEdS$, we need $\psi_\mathrm{cc} = \psi_\mathrm{c} + \left[2\pi\left(1-1/\sqrt{6}\right)+2\log{(5\cstEdS)}\right]/\sqrt{\kappa}$. The case $a>3\cstEdS$ is not physical since it is after the collapse, arising at $a=3\cstEdS$. This case features a transition from a phantom field to a real field at $a = \frac{5}{2} \cstEdS$.

Figure~\ref{fig::morphon_RZA_LCDM} shows the potential for the case ($C_2$) for $\cstEdS=500$, $\cstEdS=350$ and  $\cstEdS=280$. These values were chosen for reasons of visibility. 

In the case ($C_2$), for $a=2\cstEdS$, the potential is at a quadratic maximum, arising before the turn-around: 
\begin{eqnarray}
	\uEdS\left(a\sim2\cstEdS\right) = -\frac{5}{12\cstEdS^3}\left[ 1 + \frac{3\kappa}{4}\left(\Delta\psiEdS\right)^2\right] + \mathcal{O}\left(\Delta\psiEdS\right)^3 \, ,
\end{eqnarray}
where $\Delta\psiEdS = \psiEdS(a) - \psiEdS\left(a = 2\cstEdS\right)$. Around this maximum the morphon is a phantom field and is solution of the Klein-Gordon equation~\eref{eq::KG_phant}. Thus, as explained in 
section~\ref{sec::SFDM_SS}, we can derive an effective mass $\mEdS$ related to the potential $\UEdS$:
\begin{equation}
	\nonumber \mEdS = \hbar\initial{H} \left[\sqrt{\frac{5}{4\cstEdS^3}}\right] \ \ 
	\simeq \ \ 1.1 \, \hbar\initial{H} \, \cstEdS^{-3/2} \, . \label{eq::RZA_100_mass}
\end{equation}
This effective mass has an upper bound which is $(\mEdS)_\mathrm{max} \simeq 5.8 \, \hbar\initial{H}$, reached for $\cstEdS = \frac{1}{3}$, i.e. $\laverage{\inI}{\CM} \rightarrow -\infty$. Realistic estimates for $\mEdS$ are given in section~\ref{sec::results_SFDM}.

We derive in the next subsection the characteristics of the morphon from the RZA with a $\Lambda$CDM background, allowing to probe the influence the dark energy might have on a dark matter morphon.

\subsection{RZA Morphon---$\Lambda$CDM Background}
\label{sec::RZA_LCDM}

\begin{figure}[h]
	\centering
	\begin{tikzpicture}[scale=1.3,>=latex]
		\node (myfirstpic) at (0,0) {\includegraphics[width=\textwidth]{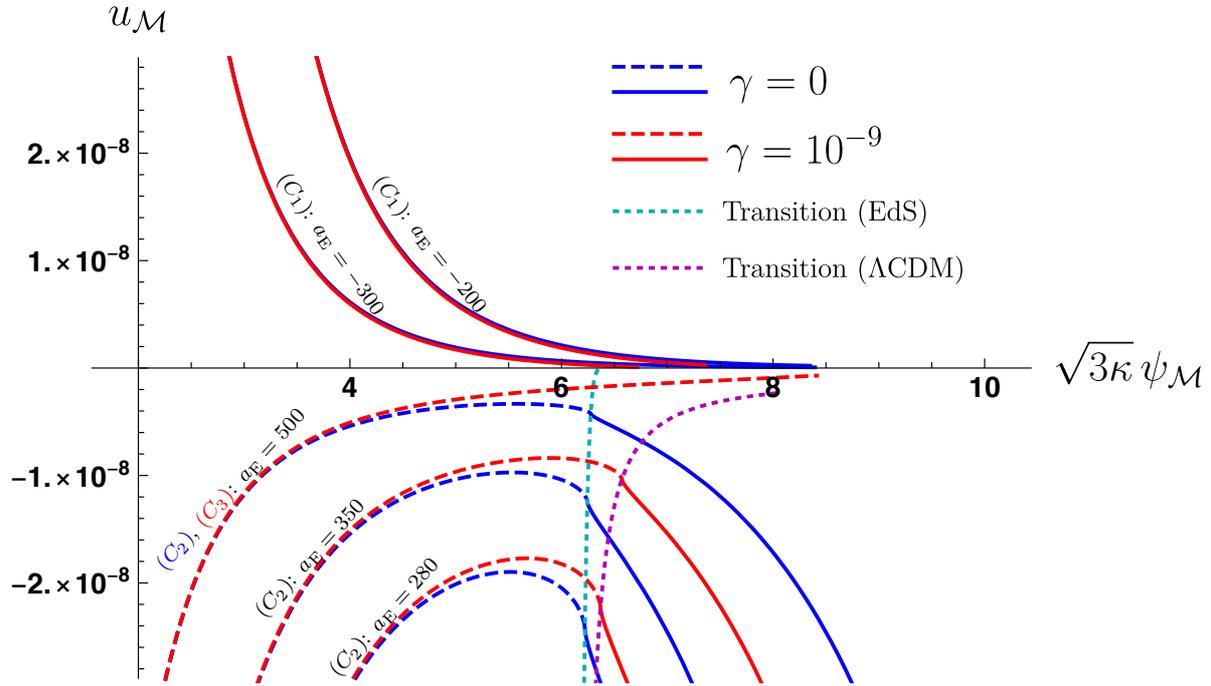}};
		\node [rotate=-52] at (-1.8,1) {\scriptsize $(C_1)$: $\cstEdS = -200$};
		\node [rotate=-55] at (-2.8,1) {\scriptsize $(C_1)$: $\cstEdS = -300$};
		\node [rotate=48] at (-3.8,-1.5) {\scriptsize \textcolor{blue}{$(C_2)$}, \textcolor{red}{$(C_3)$}: $\cstEdS = 500$};
		\node [rotate=48] at (-3.0,-2.2) {\scriptsize $(C_2)$: $\cstEdS = 350$};
		\node [rotate=48] at (-2.25,-2.8) {\scriptsize $(C_2)$: $\cstEdS = 280$};

	\end{tikzpicture}
	\caption{Adimensional potential $u_\CM$ from the curvature deviation $\CW_\CM$ as functional solution of the energy constraint in the RZA framework. The full lines correspond to a real morphon field, and the long-dashed lines to a phantom morphon field. The red curves represent the potential in the EdS case, i.e. $\gamma=0$ as calculated in section~\ref{sec::RZA_EdS}. The blue curves represent the potential for a $\Lambda$CDM background with $\gamma = 10^{-9}$ (chosen for reasons of visibility), corresponding to an initial condition at the CMB epoch and a maximum critical scale factor of $\cstLCDM^\mathrm{max} \simeq 479$. The potential is shown in both cases for different values of the critical scale factor $\cstEdS$, recalling that $\cstEdS \simeq \cstLCDM$ for $\gamma \ll 1$; from top to bottom, by pairs of curves, $\cstEdS = (-200,-300,500,350,280)$. $\psi_\CM$ is set such that $\psi_\initial{\CM} = 0$. The short-dashed light-blue curve highlights the transition from a phantom (on the left of the curve) to a real scalar field (on the right of the curve) in the ($C_2$) case for the EdS background as function of the critical scale factor; the short-dashed pink curve is the equivalent for the $\Lambda$CDM case. For $\cstEdS=500$, the blue curve is a ($C_2$) curve, but because $500>\cstLCDM^\mathrm{max}$, the red curve is a ($C_3$) curve. It never crosses the $\Lambda$CDM transition curve and then remains a phantom field, representing an ever expanding domain $\CM$.}
	\label{fig::morphon_RZA_LCDM}
\end{figure}

In the case of a $\Lambda$CDM background we have:
\begin{equation}
	a\left(t\right) = \gamma^{-1/3} \sinh^{2/3}\left(\sqrt{\frac{\gamma}{1+\gamma}}\frac{t}{\initial{t}}\right) \quad ; \quad
	\xi\left(t\right) = \frac{a \, \fnX\left(a\right) - \initial{q}}{\initial{\dot{q}}} \ , \label{eq::RZA_LCDM_xi}
\end{equation}
where $\gamma := \Omega_\Lambda^{\rm \bf i}/\Omega_\mathrm{m}^{\rm \bf i}>0$ with $\Omega_\mathrm{m}^{\rm \bf i},\Omega_\Lambda^{\rm \bf i}>0$ the initial Omega parameters of the background and the constants $\initial{q} := \fnX(1)$ and $\initial{\dot{q}} := \left[\fnX\left(1\right) + \partial_a\fnX\left(1\right)\right]\initial{H}$.\footnote{Here, $q$ represents the leading solution of the equation~\eref{eq::RZA_q} for a $\Lambda$CDM background: we have $q(t) = a \fnX\left(a\right)$.} The function $\fnX(a)$ is given by
\begin{eqnarray}
	\fnX\left(a\right) = \left(\fnG\left(a\right)\right)^{-1/3} {_2}F_1\left(\frac{1}{3}, \frac{5}{6}; \frac{11}{6}; \frac{a^3\gamma}{\fnG\left(a\right)}\right) \, ,
\end{eqnarray}
with $\fnG\left(a\right) = 1+a^3\gamma$ and $_2F_1$ is the ordinary hypergeometric function. We have the following properties:
\begin{eqnarray}
	\forall a > 1, \ \forall \gamma > 0, \quad \fnX\left(a\right) > 0 \quad \mathrm{and} \quad \fnG\left(a\right) > 0 \, ,  \\
	\forall a > 1, \  \Gamma_0\left(a\right) = 1 \quad ; \quad  X_0\left(a\right) = 1 \, , \\
	\lim_{a \to +\infty} a\fnX\left(a\right) = {_2}F_1\left(\frac{1}{3}, \frac{5}{6}; \frac{11}{6}; 1\right) \, \gamma^{-1/3} \simeq 1.44 \times \gamma^{-1/3} \, .
\end{eqnarray}
For simplicity we will write in the following  of the article $\fnX\left(a\right) = \fnX$ and $\fnG\left(a\right) = \fnG$. In the $\Lambda$CDM model, $\gamma \ll 1$ at the CMB time. Then, for $a$ such that $a^3\gamma \ll 1$, the $\xi(t)$ function is the same as in the EdS case \eref{eq::RZA_EdS_xi}. However, for $a$ corresponding to the present-epoch scale factor, dark energy is not negligible anymore and $a^3\gamma \sim 1$. At that time, the $\Lambda$CDM solutions for $a(t)$ and $\xi(t)$ diverge from the EdS case.

From equation~\eref{eq::RZA_aD} the volume scale factor $a_\CM$ is expressed as, assuming $\laverage{\inI}{\CM} \not= 0$, $\laverage{\inII}{\CM} = 0$ and $\laverage{\inIII}{\CM} = 0$:
\begin{eqnarray}
	a_\CM\left(a\right) = a \left(\frac{a \fnX - 3\cstLCDM}{\initial{q} - 3\cstLCDM}\right)^{1/3} \, , \label{eq::RZA_LCDM_aD}
\end{eqnarray}
where $\cstLCDM:=  \frac{1}{3}(\initial{q} - \initial{\dot{q}} / \laverage{\inI}{\CM})$ is a constant which can be seen as a critical scale factor.  In the limit $\gamma \ll 1$, $\initial{q} \simeq 1$ and $\initial{\dot{q}} \simeq \initial{H}$, thus $\cstLCDM \simeq \cstEdS$. Equation~\eref{eq::RZA_LCDM_aD} leads to
\begin{eqnarray}
	H_\CM\left(a\right) = \frac{\initial{H}}{3 \, a^{3/2}} \times \frac{4a\left[3\fnX\left(\fnG-\frac{1}{2}\right) + \frac{5}{2}\right] - 9\cstLCDM\fnG}{\left[a\fnX - 3\cstLCDM\right]\sqrt{\fnG\left(1+\gamma\right)}} \ . \label{eq::RZA_LCDM_HD}
\end{eqnarray}
From this expression we see that the domain $\CM$ is collapsing for $\cstLCDM \in \ ]\cstLCDM^\mathrm{min}; \cstLCDM^\mathrm{max}[$, with
\begin{eqnarray}
\fl
\cstLCDM^\mathrm{min} := \fnX(1)/3 \quad ; \quad
\cstLCDM^\mathrm{max} := {_2}F_1\left(\frac{1}{3}, \frac{5}{6}; \frac{11}{6}; 1\right) \, \frac{\gamma^{-1/3}}{3} \simeq 0.479 \times \gamma^{-1/3} \ .
\end{eqnarray}
For $\cstLCDM \notin \ [\cstLCDM^\mathrm{min} ; \cstLCDM^\mathrm{max}[$, the domain $\CM$ is ever expanding. In the limit $\gamma \ll 1$, $\cstLCDM^\mathrm{min} = 1/3$, as in the EdS case. However, the upper limit $\cstLCDM^\mathrm{max}$ has no EdS equivalent. This highlights the fact that dark energy 
dominates over small overdensities, preventing them from collapsing at late times; we recall that small overdensities correspond to small $(-\laverage{\inI}{\CM})$, i.e. high $\cstLCDM$.

From equations~\eref{eq::RZA_QD}, \eref{eq::RZA_WD} and \eref{eq::RZA_LCDM_xi} we have:
\begin{eqnarray}
	\fl \CQ_\CM = \frac{-2\initial{H}^2}{3a\fnG\left(1+\gamma\right)} \left[\frac{3\fnX - 5}{2\left(a\fnX - 3\cstLCDM\right)}\right]^2 \quad ; \quad
	\CW_\CM = \frac{-10\initial{H}^2}{a^2\left(a\fnX-3\cstLCDM\right)\left(1+\gamma\right)} \ . \label{eq::RZA_LCDM_WQ_D}
\end{eqnarray}
We introduce the adimensional, reduced potential:
\begin{equation}
\uLCDM := \frac{3\kappa}{6\initial{H}^2} U^{\Lambda}_\CM = -\frac{\mathcal{\CW_\CM}}{6\initial{H}^2} \ .
\end{equation} 
Then, from equations~\eref{eq::U_psi_dot} we obtain:
\begin{eqnarray}
	\uLCDM &=& \frac{5/3}{a^2\left(a\fnX - 3\cstLCDM\right)\left(1+\gamma\right)} \ ; \label{eq::uD_LCDM} \\
	\partial_a \psiLCDM &=& \pm\sqrt{\frac{2}{\kappa}}\sqrt{\frac{\epsilon\left[2a\left(\frac{25}{24} + \fnX\left\{\frac{5}{6}\fnG - \frac{5}{4}\right\} + \frac{3}{8}\fnX^2\right) - 5\cstLCDM\fnG\right]}{a\fnG^2\left(a\fnX-3\cstLCDM\right)^2}} \ .
	\label{eq::da_psi_LCDM}
\end{eqnarray}
Equation~\eref{eq::da_psi_LCDM} is not analytically integrable. We can, however, infer the behaviour of the potential as a function of the parameter $\cstLCDM$, as in section~\ref{sec::RZA_EdS}. Three cases can be distinguished: the first two are equivalent to the EdS solution ($C_1$) with $\cstLCDM < \cstLCDM^\mathrm{min}$, the domain $\CM$ is underdense with respect to the background and is ever expanding (not considered here), with a real morphon field; ($C_2$) $\cstLCDM^\mathrm{min} < \cstLCDM < \cstLCDM^\mathrm{max}$, the domain $\CM$ is overdense and collapsing with a morphon featuring a transition from a phantom to a real field. The third case is particular to the $\Lambda$CDM background and reflects the fact that dark energy can become dominant over small overdensities at late times: ($C_3$), the domain $\CM$ is overdense and ever expanding with a phantom morphon field. These solutions are summarized in figure~\ref{fig::morphon_RZA_LCDM}, representing the adimensional potential $u_\CM$ as a function of the scalar field $\psi_\CM$ for the two background models.

As for the EdS background, we can define an effective mass $\mLCDM$ for the phantom morphon field in the case ($C_2$). However, this mass is not analytical for any $\cstLCDM$, but we can derive it for $\cstLCDM=\cstLCDM^\mathrm{min}$, leading to a maximum effective mass $\left(m^\mathrm{\Lambda}_{\rm eff}\right)_\mathrm{max}$, and for $\cstLCDM=\cstLCDM^\mathrm{max}$, leading to a minimum effective mass 
$(\mLCDM)_\mathrm{min}$. We have:
\begin{subequations}
	\label{eq::RZA_100_LCDM_mass}
	\begin{eqnarray}
	(\mLCDM)_\mathrm{min} &=& \hbar\initial{H}\sqrt{\frac{2\gamma}{1+\gamma}} = \hbar \sqrt{\frac{2\Lambda}{3}}, \quad \forall \gamma \, ; \\
	(\mLCDM)_\mathrm{max} &=& \hbar\initial{H}\sqrt{\frac{135}{4}} = (\mEdS)_\mathrm{max} \, , \quad \mathrm{for} \quad \gamma \ll 1 \, .
	\end{eqnarray}
\end{subequations}
We see that $(\mLCDM)_\mathrm{min}$ does not depend on the initial Hubble parameter, i.e. on the initial time $\initial{t}$. With $\Lambda = 1.11 \times 10^{-52}$~m$^{-2}$ \cite{Planck2018VI}, we get $(\mLCDM)_\mathrm{min} \sim 10^{-52}$~eV. At the maximum of the potential, where $\mLCDM$ is defined, the curvature is non-zero, and it is of the same order of magnitude as in the EdS case. Realistic estimates for $\mLCDM$ are given in section~\ref{sec::results_SFDM}.

\subsection{RZA and Scaling Solution}
\label{sec::RZA_SS}

In this subsection we make use of the upper-scripts $\rm SS$ and $\rm RZA$, to refer to the common quantities used in the scaling solution models and in the RZA models.

The set of initial conditions differs as a function of the model considered. For the scaling models we have to set $\{ H^{\rm SS}_\initial{\CM}; \gamma^{\initial\CM}_{\CW\CQ\mathrm{m}}; n \}$; for the RZA models we have to set $\{ \initial H; \laverage{\inI}{\CM}^{\rm RZA}; \gamma\}$. The initial averaged density $\laverage{\varrho}{\CM}^{\rm RZA}$ is given by $\laverage{\varrho}{\CM}^{\rm RZA} = \initial\varrho\left(1-\laverage{\inI}{\CM}^{\rm RZA}/\initial{H}\right)$, where $\initial\varrho$ is the initial matter density of the background. In this subsection we will only use the RZA-EdS solution, then $\gamma = 0$.

In order to be able to compare these two models, we must have the same set of initial parameters for both of them. To this end, we assume the collapsing geometry of the scaling solution models to be the same as in  our RZA models, i.e. a strongly anisotropic (\textit{locally} plane-symmetric) collapse (see \ref{appA}). Then, we take $\laverage{\inII}{\CM}^{\rm SS} = 0$, $\laverage{\inIII}{\CM}^{\rm SS} = 0$ and $\laverage{\inI}{\CM}^{\rm SS} = \laverage{\inI}{\CM}^{\rm RZA}$. We only use the critical scale factor $\cstEdS$ to refer to the averaged first invariant. This assumption leads to the following relation:
\begin{eqnarray}
	\gamma^{\initial\CM}_{\CW\CQ\mathrm{m}}\left(\cstEdS, n\right) = -\left(\frac{9}{4}\left(n+2\right)\left(1-3\cstEdS\right)^2\left(\frac{H^{\rm SS}_\initial{\CM}}{\initial H}\right)^2 + 1\right)^{-1} \, . \label{eq::gamma_SS_RZA}
\end{eqnarray}
We are left with two free parameters for the scaling solution, $H^{\rm SS}_\initial{\CM}$ and $n$.

A first choice is to take the same initial conditions for both models, i.e. the same initial \textit{domain-dependent} expansion rates: with $H^{\rm SS}_\initial{\CM} = H^{\rm RZA}_\initial{\CM}$. Thus, the domain $\CM$ in both cases has a lower initial expansion rate than the global model universe. The only free parameter remaining, compared to the RZA models, is the scaling exponent $n$. As studied in \cite{Wiegand2010}, the scaling solution allows us to describe the evolution of large-scale subdomains of the Universe, the solution with $n=-1$ being the leading-order perturbative solution to the general cosmological equations. This argument cannot be used in our case because we consider galaxy cluster scales. However, one advantage of the $n=-1$ solution is to lead to a positive curvature which is in agreement with the RZA solutions and with what is expected from overdense regions of the Universe. A second advantage is linked to the expression of the domain scale factor at turn-around,~i.e.~$a_{\CM_{\rm ta}}$. In the RZA model, with an EdS background, the domain-dependent scale factor at turn-around is given by:
\begin{eqnarray}
	a_{\CM_{\rm ta}}^{\rm RZA} = \frac{9}{4} \cstEdS \left(\frac{-\frac{3}{4} \cstEdS}{1-3\cstEdS}\right)^{1/3} \, .
\end{eqnarray}
For a small initial overdensity, i.e. $-\laverage{\inI}{\CM}^{\rm RZA}/\initial{H} \ll 1$, or $\cstEdS \gg 1$, we have $a_{\CM_{\rm ta}}^{\rm RZA}~\sim~2^{-5/3} \times \frac{9}{2} \cstEdS$. In the scaling solution, it is given by (section~\ref{sec::SS}):
\begin{eqnarray}
a_{\CM_{\rm ta}}^{\rm SS} = \left(\frac{1}{4}\left(n+2\right)\left(4-9\cstEdS\right)^2 + 1\right)^{\frac{1}{n+3}} \, .
\end{eqnarray}
For $\cstEdS \gg 1$, we have $a_{\CM_{\rm ta}}^{\rm SS} \sim \frac{9}{2} \cstEdS^{2/(n+3)}$. We infer that the value $n=-1$ implies a similar turn-around domain scale factor for small overdensities, with $a_{\CM_{\rm ta}} \propto \cstEdS$ for both class of models. This speaks in favour of the $n=-1$ scaling solution.

However, the two turn-around domain scale factors are not equal in the limit $\cstEdS \gg 1$, and we have $a_{\CM_{\rm ta}}^{\rm SS} \simeq 3.2 \times a_{\CM_{\rm ta}}^{\rm RZA}$. The same remark can be made for the turn-around time with $t_{\rm ta}^{\rm SS} \not= t_{\rm ta}^{\rm RZA}$. Thus, the volume of the domain at turn-around is bigger in the case of the scaling solution, implying a later collapse. This shows that this class of solutions is not really appropriate for collapsing subdomains after turn-around.

The second choice to set the free parameters $H^{\rm SS}_\initial{\CM}$ and $n$ is to fit the turn-around of the RZA solution with the scaling solution, i.e. assuming $t_{\rm ta}^{\rm SS} = t_{\rm ta}^{\rm RZA}\left(\cstEdS, \initial H\right)$ and $a_{\CM_{\rm ta}}^{\rm SS} = a_{\CM_{\rm ta}}^{\rm RZA}\left(\cstEdS\right)$. This method is detailed in \ref{appB}. It leads to two solutions:
\begin{itemize}
	\item In the first solution, $H^{\rm SS}_\initial{\CM}\left(\cstEdS\right) > \initial H$. The corresponding scaling exponent $n\left(\cstEdS\right)$ tends to the value $n = -2$ from above in the limit $\cstEdS \rightarrow +\infty$. This limit is reached quickly with $n=-1.9$ already for $\cstEdS = 2$. For realistic values of $\cstEdS$ (given in table~\ref{tab::mass}), we can consider $n\simeq -2$.
	\item In the second solution, $H^{\rm SS}_\initial{\CM}\left(\cstEdS\right) < \initial H$. The scaling exponent $n\left(\cstEdS\right)$ tends to the value $-1$ from above in the limit $\cstEdS \rightarrow +\infty$. The convergence is very slow: for realistic values given in table~\ref{tab::mass} we have $n(26) \simeq -0.3$ and $n(236) \simeq -0.6$.
\end{itemize}
Both solutions well fit the RZA-EdS solution until the turn-around. As expected for reasons presented previously, they diverge after the turn-around, with a later collapse for the scaling solutions. However, the second solution, i.e. with $H^{\rm SS}_\initial{\CM}\left(\cstEdS\right) < \initial H$, is more representative of our collapsing domain, as we expect the initial domain expansion rate to be smaller than in the EdS solution. This justifies the value $n \simeq -1$ taken in the first choice made in this subsection.

\begin{table}[h]
\centering
	\caption{Effective mass of the morphon field for the scaling solution, the RZA$_{\rm EdS}$ and the RZA$_{\rm \Lambda CDM}$ models. The values of the averaged initial first invariant depends on the size of the domain $\CM$. They are chosen to be at 1$\sigma$ of density contrast for each domain size, as presented in section~\ref{sec::IC}. The initial condition is at redshift $200$. The size of the domain is given at present time supposing a linear evolution of the domain, hence $R_0 = \frac{q_0}{\initial{q}} \initial R$ where $q(t)$ is defined in equation~\eref{eq::RZA_q} \cite{Bildhauer1992}.}
	\vspace{.3cm}
	\begin{tabular}{l r c }
		\midrule \midrule
		$\frac{q_0}{\initial{q}} \initial{R}$ & 5 Mpc &  50 Mpc \vspace{.05cm}  \\ 
		\vspace{.05cm}  $\laverage{\inI}{D}/\initial{H}$ [at $1\sigma$] & $13 \times 10^{-3}$ & $1.4 \times 10^{-3}$ \vspace{0.01cm} \vspace{.05cm} \\
		\vspace{.05cm}  $\cstEdS \simeq \cstLCDM$ & $26$ & $238$ \vspace{0.01cm} \vspace{.05cm} \\
		\hline\hline \vspace{-.4cm} \\
		& \multicolumn{2}{c}{ ${ m_{\rm eff} \,   [eV] }$} \\
		& \multicolumn{2}{c}{ $\overbrace{\hspace{5cm}}$} \\
		SS (n=-1) & $5.2 \times 10^{-33}$ & $1.8 \times 10^{-34}$ \\
		RZA$_{\rm EdS}$ & $3.4 \times 10^{-32}$ & $1.2 \times 10^{-33}$ \\
		RZA$_{\rm \Lambda CDM}$ & $1.8 \times 10^{-32}$ & No Collapse \\
		\midrule \midrule
	\end{tabular}
	\label{tab::mass}
\end{table}

\subsection{Initial conditions}
\label{sec::IC}

Initial data are given at redshift $z=200$, allowing us to be at dust-dominated epochs but still in the linear regime. The initial Hubble parameter is $\initial{H}^{\rm EdS} \simeq 1.9 \times 10^5 \rm \; km \, s^{-1} \, Mpc^{-1}$ for an EdS background, and $\initial{H}^{\rm \Lambda CDM} \simeq 1.0 \times 10^5 \rm \; km \, s^{-1} \, Mpc^{-1}$  for a $\Lambda$CDM background. The initial averaged first invariant is taken from \cite{BKS2000} to be at 1$\sigma$ of the initial perturbation fluctuation (Table~1 in \cite{BKS2000} that gives values for the dimensionless averaged invariant $\laverage{\inI}{\CM}/\initial H$ for different smoothing scales). As outlined in section~\ref{sec::TheRZA}, we are interested in scales from 5 Mpc up to 50 Mpc.
In section~\ref{sec::results_power_spectrum}, a discussion is made on the dependence of the averaged first invariant in \cite{BKS2000} with the amount of dark matter considered in the initial power spectrum.

\section{Results}
\label{sec::results}

Our results for the two approaches are discussed by firstly comparing our approach to SFDM models, where we provide the effective mass expected from the morphon analogy. Secondly, the outcome of the RZA model is discussed focussing on the role of average intrinsic curvature and of kinematical backreaction.

\subsection{Morphon field and SFDM}
\label{sec::results_SFDM}

We derived the morphon field from exact and approximate solutions of the general cosmological 
equations~\eref{eq::Buchert_eff} for an irrotational dust fluid, firstly with the exact class of scaling solutions in section~\ref{sec::Morph_SS}, secondly, with the RZA at an EdS background in section~\ref{sec::RZA_EdS}, and, thirdly, with the RZA at a $\Lambda$CDM background in section~\ref{sec::RZA_LCDM}.
As we are only interested in collapsing overdense domains of the Universe for this study, we only consider morphons resulting from such domains. We expect them to have a positive average scalar curvature. The corresponding morphons, for each presented solution, all feature a phantom field, either totally as in the scaling solution (equation~\eref{eq::phi_U_sin_bis} with $\epsilon = -1$), or partially as in the RZA solution (see figure~\ref{fig::morphon_RZA_LCDM}). This result is radically different from the type of scalar fields used in the SFDM models where the fundamental fields are always real (unless more general scalar field theories or special hypotheses are invoked that allow for crossing the phantom divide, e.g. \cite{Alimi}). One of the reasons is that SFDM models used for cosmological dark matter in a dynamical description are global models for the whole model universe, whereas we stick to a regional description of a collapsing overdense subdomain. Our collapse condition leads to a phantom field part.

Even if we cannot directly compare our effective potentials with potentials used in SFDM models, we are still able to define an effective mass $m_{\rm eff}$ for the phantom scalar field (see section~\ref{sec::SFDM_SS}). For the scaling solution it is given by equation~\eref{eq::SS_mass}, for the RZA it is fully analytic only for an EdS background, equation~\eref{eq::RZA_100_mass}. In a $\Lambda$CDM background we can only derive the maximum range of possible values for the effective mass (equation~\eref{eq::RZA_100_LCDM_mass}), the intermediate values must be numerically integrated.

In table~\ref{tab::mass} we give the values of $m_{\rm eff}$ for each solution and for two different spatial scales. The initial conditions are taken from section~\ref{sec::IC}. The effective mass $m_{\rm eff}$ is smaller in the case of the scaling solution. However, this model served as an illustration; these kind of models are not realistic for relatively small domains where non-linearities play an important role in the collapsing process. They provide, however, efficient solutions for larger scales as shown in \cite{Wiegand2010}. The effective masses in both RZA models are of the same order of magnitude, due to the small changes expected in the shape of the potential between EdS and $\Lambda$CDM. Compared to typical values used in SFDM models for the mass of the field, around $10^{-22}$ eV, the values in table~\ref{tab::mass} are about $10$ orders of magnitude smaller. This is yet another difference to fundamental fields. \\

\subsection{Curvature and Dark Matter}
\label{sec::results_Omegas}

In this subsection we focus on the RZA model and do not refer to the scaling solution as it does not properly model nonlinearities that are essential for the final evolution of the collapsing domain.

As explained in section~\ref{sec::TheRZA}, the functionals used in RZA depend on which of the equations in the set of the $3+1$ Einstein equations is employed to derive them. Aiming at conserving certain equations implies the violation of others due to the fact that RZA only provides an approximation. In this article we use the continuity equation (for $a_\CM$), the definition of $\CQ_\CM$ in terms of the averaged scalar invariants, and a combination between the averaged energy constraint and the averaged Raychaudhuri equation (for $\CW_\CM$). It turns out that with this choice, along with the vanishing of averaged second and third invariants, the energy constraint is on average preserved. Thus, it makes sense to study the $\Omega$-parameters to determine which component is dominating in the energy budget.

\begin{figure}[h!]
\centering
\includegraphics[width=\textwidth]{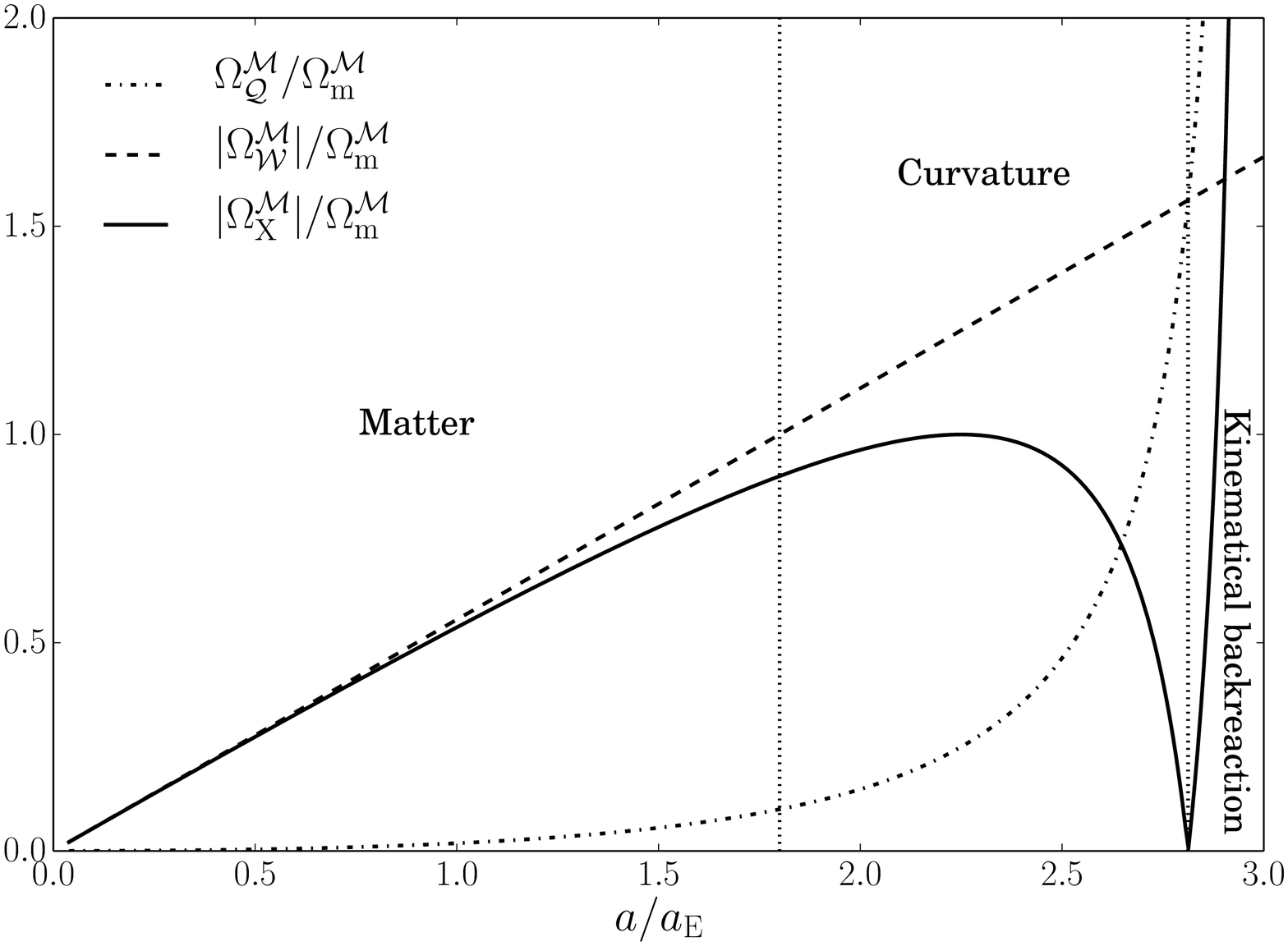}
\caption{Ratios $\Omega_\CQ^\CM/\Omega_{\rm m}^\CM$ (dashed-dotted line), $\left|\Omega_\CW^\CM\right|/\Omega_{\rm m}^\CM$  (long-dashed line), and $\left|\Omega_{\rm X}^\CM\right|/\Omega_{\rm m}^\CM$ (full line) as functions of the rescaled background scale factor $a/a_{\rm E}$  for the RZA model with an EdS background. The turn-around is reached at $a/a_{\rm E} = 2.25$. The dominating components are indicated; the dominance zones are delimited by vertical short-dashed lines. The dominating component is initially matter, then the curvature before and after the turn-around, and finally the kinematical backreaction until the end of the collapse. The figure holds on all considered time-scales because of the rescaled x-axis.}
\label{fig::Omegas}
\end{figure}

In figure~\ref{fig::Omegas} we show, until the collapse time, the ratios $\Omega_\CQ^\CM/\Omega_{\rm m}^\CM$, $\left|\Omega_\CW^\CM\right|/\Omega_{\rm m}^\CM $, and $\left|\Omega_{\rm X}^\CM\right|/\Omega_{\rm m}^\CM$ as functions of the rescaled background scale factor $a/a_{\rm E}$ in an EdS background (rescaled to render the figure independent of the absolute collapse time). The figure is also independent of the scale and mass of the domain $\CM$ because these properties are fully described by the parameter $\cstEdS$. $\Omega_{\rm X}^\CM$ comprises the total deviation from the standard model due to backreaction terms and is defined as $\Omega_{\rm X}^\CM := \Omega_\CQ^\CM + \Omega_\CW^\CM$. The figure shows that the curvature cannot be neglected in the evolution of the domain. Initially, the dust matter dominates over the two other components. The curvature becomes dominant before the turn-around, arising at $a/\cstEdS = 2.25$. At the end of the collapse, the kinematical backreaction dominates over the matter and the curvature. We see that both the kinematical backreaction and the curvature cannot be neglected at all times during the collapse process. The Relativistic Zel'dovich Approximation shows that the curvature is the component driving the dynamics for most of the evolution. The kinematical backreaction is non-negligible only at the end of the collapse, which should then lead to virialization of the domain. This latter cannot be described with the present model, since other physical effects become relevant, such as velocity dispersion, baryonic pressure and vorticity. Since the model is limited to the pre-virialization regime, it cannot predict the actual fraction of `dark matter' $\Omega^\CM_{\rm X}$ (here due to backreaction) and `fundamental energies' contained in $\Omega_{\rm m}^\CM$. However, figure~\ref{fig::Omegas} shows that the `dark matter' already contained in the curvature provides about twice the energy contained in $\Omega_{\rm m}^\CM$ at the onset of virialization; $\Omega^\CM_{\rm X}$ becomes dominated by the kinematical backreaction (averaged rate of shear) only at the end of the collapse (at that time $\Omega^\CM_{\rm X}<0$), but is then in competition with pressure effects and vorticity.

The contribution of backreaction, depicted by $\Omega^\CM_{\rm X}$, is subdominant with respect to $\Omega^\CM_{\rm m}$ before the turn-around, but then quickly reaches the same order. A striking result of our analysis is that already at the turn-around the total contribution of the inhomogeneity effects becomes equal to the contribution of the fundamental matter source. This is indeed expected, since at turn-around we have $H_{\CM_{\rm ta}} = 0$  and, thus, $2\kappa \langle\varrho\rangle_{\CM_{\rm ta}} = (\CQ_{\CM_{\rm ta}} + \CW_{\CM_{\rm ta}})$.

Finally, a remark is in order as to the question of disentangling general relativistic backreaction models from Newtonian backreaction models. Although Newtonian simulations assume zero curvature, they describe the collapse process realistically in terms of density and kinematical variables like expansion, shear, vorticity and velocity dispersion. Hence, kinematical backreaction is present on the scale of our study \cite{BKS2000}. In the Newtonian backreaction model we can also set up an averaged energy balance equation where the average curvature is replaced by
a term that contains the time-history of backreaction, being obtained via formal integration of the Newtonian volume acceleration law (or through integration of the integrability condition \eref{eq::integrability} in general relativity). We obtain \cite[Sect. 2.3.1 for $\CW_\CM$, and Sect. 2.4.1 for $\Omega^\CM_{\rm X}$]{Buchert2008}:
\begin{equation}
\label{newtoncurvature}
\CQ_\CM + \CW_\CM = - \frac{2}{a_\CM^2} \int_{\initial t}^{t} {\mathrm d}t' \, \CQ_\CM \frac{\mathrm d}{{\mathrm d}t'} 
a^2_\CM (t') \ .
\end{equation}
The question of disentanglement is not answered in this paper, since we confine ourselves to the general relativistic context. It will be interesting to go deeper into this question, especially addressing the following 
two points: (i) investigating the average behaviour of a local realization of the Newtonian Lagrangian scheme, focussing on whether it can reproduce a curvature-like term, \eref{newtoncurvature}, predicted in the framework of the average dynamics, and (ii) comparing the average curvature from the averaged energy constraint with the average curvature resulting from the RZA metric and its derivatives \cite{RZA_1}; for point (ii) we know that in the Lagrangian linearized regime both match by construction, but they may deviate from each other in the nonlinear regime of the collapse.

Even if a Newtonian simulation can reproduce a `curvature-like term', dark matter estimates are often done in simplified (spherically symmetric and sometimes static) models where kinematical backreaction and its time-history vanish. 

\subsection{Issues of dark matter \textit{vs.} general relativity}
\label{sec::results_power_spectrum}

The estimates made in \cite{BKS2000}, used in section~\ref{sec::results_SFDM}, assumed a dark matter-dominated initial power spectrum (Appendix~G in \cite{Bardeen1986}). Since we seek to probe the potential dark matter behaviour of the morphon, with respect to SFDM models, we should use a dark matter-free initial power spectrum. For sufficiently large scales, we assume the initial power-law dark matter-free spectrum to be just, on average, rescaled by a factor. This would imply a change of the averaged first invariant by this same factor to smaller amplitudes. Accordingly, we expect $\cstEdS$ to be bigger with a maximum change of one order of magnitude. This does not significantly change the results of section~\ref{sec::results_SFDM}.
The different clustering properties of baryonic matter compared with dark matter is not an issue, since we restrict the collapse model to megaparsec scales, where pressure and other effects in baryonic matter do not play a role.

For section~\ref{sec::results_Omegas}, The different phases of dominance for each component are independent of $\cstEdS$. Only the timing of the collapse changes with a rescaling of the initial power spectrum, which is eliminated by the choice of the x-axis in figure~\ref{fig::Omegas}. A detailed statistical analysis of the abundance of collapsed objects as a function of time lies beyond the scope of this article. We do not seek to conduct an extensive study of the mass function of collapsed objects that would be necessary for a quantitative investigation of the backreaction distributions for a dark matter-free power spectrum.\footnote{For the effect of backreaction on the mass function, see \cite{bolejkojan}.}
 
As presented in the introduction, the term \textit{dark matter} and its postulated fundamental origin results from different phenomena which are currently best explained by some kind of massive particle that has weak interactions with baryonic matter. Those phenomena are: the relative height of the acoustic peaks in the power spectrum of CMB (Cosmic Microwave Background) anisotropies; the energy budget of the different energy components of the Universe acting on its global expansion;  the timing of collapse of large-scale structures in relation to the initial amplitude of the power spectrum and the epoch of structure formation; the virialization of clusters of galaxies related to the energy budget in large-scale structures on megaparsec-scales; the flat rotation curves of galaxies. Handling all these issues in a novel approach, by a general-relativistic modeling of structure, needs to be built up in steps. In this article we only focus on the energy budget needed in the collapse of structures on megaparsec-scales.

Some remarks are in order to recall the often implicit model assumptions made that lead to the conclusion that dark matter is needed. In current simulations of structure formation it is assumed that a single FLRW metric describes the average on all scales except in the vicinity of strong-field objects \cite{ishibashiwald}. While it has been acknowledged that the argument of small metric perturbations even for large masses is not sufficient to justify an all-scale FLRW metric since the derivatives of the metric, especially its second derivatives (curvature), can be large \cite{gw},\footnote{See also\cite{curvatureestimates} for estimates of curvature in observed structures.} the hypothesis that structure averages out on an assumed global background is still held and implemented. This is true for Newtonian simulations but also for general relativity-based simulations, which suppress nonlinear curvature evolution due to a global background architecture, e.g. \cite{Adamek}. However, there have been a number of clear demonstrations that this assumption is too restrictive. It even runs intro contradictions as was recently shown by comparing the approach of \cite{gw} with the approach employed in this paper \cite{cliftonsussman}.\footnote{For a full assessment of the backreaction problem in comparison with the approach by \cite{gw} see \cite{buchert11}. Note that there are also other issues of coarse-graining and clock synchronization beyond the classical implementation of general relativity \cite{Wiltshire:weakfield,NazerWiltshire,visser}.}
Further demonstrations in terms of exact solutions show that general relativity leads to strong curvature effects, even if strong-field sources are negligible in terms of volume measure as compared to the total volume of a given spatial domain \cite{Mikolaj1}, and even if the hierarchy of structures is well-described as a `weak field' perturbation \cite{Rasanen:weakfield} at each level in the hierarchy \cite{Mikolaj2}.

The assumption of a global background is a key ingredient of arguments connecting the different dark matter issues outlined above. That we cannot compare structures with respect to such a global background is essentially due to (i) non-commutation of averaging and time-evolution that is different on different scales \cite{ellisbuchert}, and (ii) non-existence of a conservation law for curvature, i.e. the fact that positive curvature in overdense regions and negative curvature in void regions do not average out to a globally assumed background curvature; instead a combined conservation law exists that couples the average curvature to structure fluctuations in a scale-dependent way \cite{Buchert2000,BuchertCarfora2008}.
 
\section{Conclusions} \label{sec::discussion}

We analyzed exact and approximate solutions of the averaged Einstein equations in order to understand to which extent inhomogeneities can mimic dark matter. This question has been investigated using two approaches: firstly, by comparing the effective scalar field (the morphon field) resulting from inhomogeneities to fundamental scalar fields used in Scalar Field Dark Matter (SFDM) models; secondly, by comparing the influence of the kinematical backreaction and the averaged scalar curvature during the collapse of $5-50$ Mpc overdense unions of subdomains $\CM$ of the Universe. Two types of solutions have been used for this study: exact scaling solutions in section~\ref{sec::Morph_SS}, and an approximate solution using a relativistic Lagrangian perturbation approach in section~\ref{sec::RZA}. The former cannot be used to accurately describe structure formation because of the failure to reliably model nonlinear effects. It, however, reproduces well qualitative results needed for the description of a collapsing overdense domain: the averaged curvature is positive, and the morphon field is mainly a phantom field. The second solution allows for a more realistic description of structure formation. While we chose for this article the restricted case $\laverage{\inI}{\CM} \not= 0$, $\laverage{\inII}{\CM} = 0$ and $\laverage{\inIII}{\CM} = 0$, we argue in \ref{sec::TheRZA} and in \ref{appA} with heuristic arguments that this is a reasonable choice for the description of collapsing overdense megaparsec-sized regions of the Universe before shell-crossing (see also the related investigation through generic realizations of RZA in \cite{JanBoud}).

The morphon fields derived for each model, in the collapsing case, all feature a phantom part for most part of the potential.
We recall that in this effective approach there is no violation of energy conditions. Assuming $1\sigma$ (typical) initial conditions, the effective masses derived from these potentials are of the order of 10$^{-32}$ eV to 10$^{-33}$ eV. This does not reproduce the results of SFDM models. This can be in part traced back to the fact that SFDM models consider domains representative of the Universe, whereas we focus on collapsing subdomains, where dark matter plays a dominant role in the process of collapse. Of course, SFDM models are justified only on phenomenological grounds, while the collapse models studied can be justified from physical assumptions.

We illustrated with a realistic model, in a restricted case, that during the collapse phase the average scalar curvature is positive. Along with the kinematical backreaction, they both subsequently dominate over the matter content on the domain $\CM$, the curvature being the main driver of the collapse. Recall, section~\ref{sec::results_Omegas}, that kinematical backreaction is also present in a Newtonian simulation of the collapse process \cite{BKS2000}, but it here couples to the average scalar curvature. 
Curvature is not a `relativistic correction' but an equal player in the energy budget. 
Our results imply that inhomogeneities, essentially through curvature, may play a major role in the formation of clusters of galaxies, and they have large impact on the amount of \textit{cosmological dark matter} needed in the energy budget of a collapsing overdense megaparsec-scale region of the Universe.

\ack \vspace{-10pt}
This work is part of a project that has received funding from the European Research Council (ERC)
under the European Union's Horizon 2020 research and innovation programme (grant agreement ERC advanced grant 740021--ARTHUS, PI: TB). QV is supported by a `sp\'ecifique Normalien' Ph.D. grant from the 
\'Ecole Normale Sup\'erieure de Lyon. We thank Jean-Michel Alimi and \'Etienne Jaupart for valuable discussions, Pierre Mourier for his constant interest and detailed comments on the manuscript, and Jan Ostrowski for several useful discussions on the RZA framework. We also like to thank Paul Godart who studied the morphon analogy with dark matter during a master internship in 2016 for his insights (internship report available on request \cite{Godart2016}). We are thankful to the anonymous referees for their thorough evaluations.

\appendix

\section{Dominance of the first principal scalar invariant of the expansion tensor}
\label{appA}

In order to analytically illustrate the expected form of the morphon potential from a Lagrangian perturbation approach, we have neglected the initial second and third principal scalar invariants of the expansion tensor in the expressions of the backreaction functionals, evaluated in the main text. Although a full-scale investigation of this approach is possible, we have limited our considerations to a case that allows for a transparent construction of analytical expressions. Below, we give arguments in support of this simplifying choice to show that it nevertheless captures important features and the leading terms of a collapsing domain.

\subsection{Local argument}

A basic insight underlying the power of the classical Zel'dovich approximation to describe a collapsing volume element 
is the feature of a maximally anisotropic collapse, see e.g. \cite{zeldovich,doroshkevich,szalay}. This insight is \textit{local} and simply follows from the consideration of the eigenvalues of the expansion tensor, let us call them $\lambda_1$, $\lambda_2$ and $\lambda_3$ and let us order them such that $\lambda_1$ is initially larger than the others, and that we look at the sign corresponding to the collapsing situation (the expansion tensor is locally diagonalizable for irrotational flows, but the eigenvalues are functions of the Lagrangian coordinates).
Since the Lagrangian perturbation solutions are separable into initial data and a global time-function (here $\xi (t)$), these eigenvalues are multiplied by time-functions in the course of evolution such that the difference between the eigenvalues increases, leading to the dominance of $\lambda_1$ before the collapse. Hence, the principal scalar invariants expressed in terms of the dimensionless eigenvalue functions, $\inI = \lambda_1 + \lambda_2 + \lambda_3$, $\inII = \lambda_1 \lambda_2 + \lambda_1 \lambda_3 + \lambda_2 \lambda_3$,
$\inIII = \lambda_1 \lambda_2 \lambda_3$, are expected to be dominated by the first invariant $\inI \approx \lambda_1$, $\inII \approx 0$, $\inIII \approx 0$. The second and third invariants at later times vanish in the RZA, if their initial data vanish.

\subsection{Regional argument}

However, we employ the average properties of RZA which leads us to the need to extend the argument above to a regional collapsing domain. For this purpose, we are going to investigate the following geometric toy model that can be seen as a model illustrating the well-known Lin-Mestel-Shu instability of a spherically symmetric collapse \cite{linmestelshu}.

We restrict ourselves to the Newtonian case to study the kinematical properties of a collapsing spheroid, and we
employ the integral-geometric formulae for the backreaction functional, \cite[Sect. 3.1.2, Eq. (58)]{Buchert2008}.
We calculate the generalization of the formulae (59) in \cite{Buchert2008} for the Minkowski functionals \cite{Mecke} of a ball to the Minkowski functionals $W_{\alpha}, \alpha = 0,1,2,3$, of an oblate spheroid with major semi-axis $a$ and excentricity $e$
($W_3$ is related to the Euler characteristic which is not needed for the backreaction functional):
\begin{eqnarray}
		W_0 (s) = \frac{4\pi}{3} a^3 \sqrt{1-e^2} \ ; \label{eq::W0_AN} \\
		W_1 (s) = \frac{4\pi}{3} a^2 \left[ \frac{1}{2} + \frac{1-e^2}{2e} \, \mathrm{artanh}\left(e\right)\right] \, ; \\
		W_2 (s) = \frac{4\pi}{3} a \left[\frac{\sqrt{1-e^2}}{2} + \frac{1}{2e}\mathrm{arctan}\left(\frac{e}{\sqrt{1-e^2}}\right)\right] \, ; \\
		W_3 (s) = \frac{4\pi}{3} \ ,
\end{eqnarray}
where the functionals are evaluated on the boundary $\partial\CM$ of the averaging domain, $s = \rm const$.
The averaged invariants and the kinematical backreaction are given by:\footnote{\textit{Erratum:}: In 
\cite[Sect. 3.1.2]{Buchert2008} there are typos in equations (55) and (56), where the division by the volume is missing in the second equalities; in equation (56), there should also be a factor of $1/3$ in front of the integrated Gaussian curvature from its definition.}
\begin{eqnarray}
	\average{\inI}{\CM} (s) = 3 \frac{W_1}{W_0}  \quad ; \quad \average{\inII}{\CM} (s) = 3 \frac{W_2}{W_0} \quad ; \quad \average{\inIII}{\CM} (s) = \frac{W_3}{W_0} \quad ;  \\ 
	\CQ_\CM (s) = 2\average{\inII}{\CM} - \frac{2}{3} \average{\inI}{\CM}^2 \, =\, 6 \left(\frac{W_2}{W_0} - \frac{W_1^2}{W_0^2}\right)  \, , 
	\label{eq::I_II_AN}
\end{eqnarray}
where the averaged third invariant is related to the Euler characteristic $\chi$ of the domain, $\average{\inIII}{\CM} W_0= \frac{4\pi}{3}\chi $ (an integral of motion in terms of the parameter $s$ for regular solutions; $\chi =1$ for simply-connected domains) \cite[Sect. 3.1.2, Eq. (57)]{Buchert2008}, and where on a spherical region $\CM = \CB$ the backreaction vanishes, $\CQ_\CB =0$ providing the integral-geometric proof of Newton's iron sphere theorem \cite[Sect. 3.1.2, Eq. (59)]{Buchert2008}.
 
The domain is initially quasi-spherical, with an eccentricity close to zero. During its collapse, the eccentricity will increase until reaching a value close to $1$ at the time before shell-crossing will occur (i.e. the pancake singularity). At this time the normal direction to the collapse will stop being degenerate. A lateral collapse with respect to the axis of revolution of the spheroid will appear, leading to a filament and finally a stable cluster in the final stage of the evolution. In this article we are only interested in the first phase of this evolution, i.e. before the first shell-crossing.

\begin{figure}[h]
\centering
\includegraphics[width=0.8\textwidth]{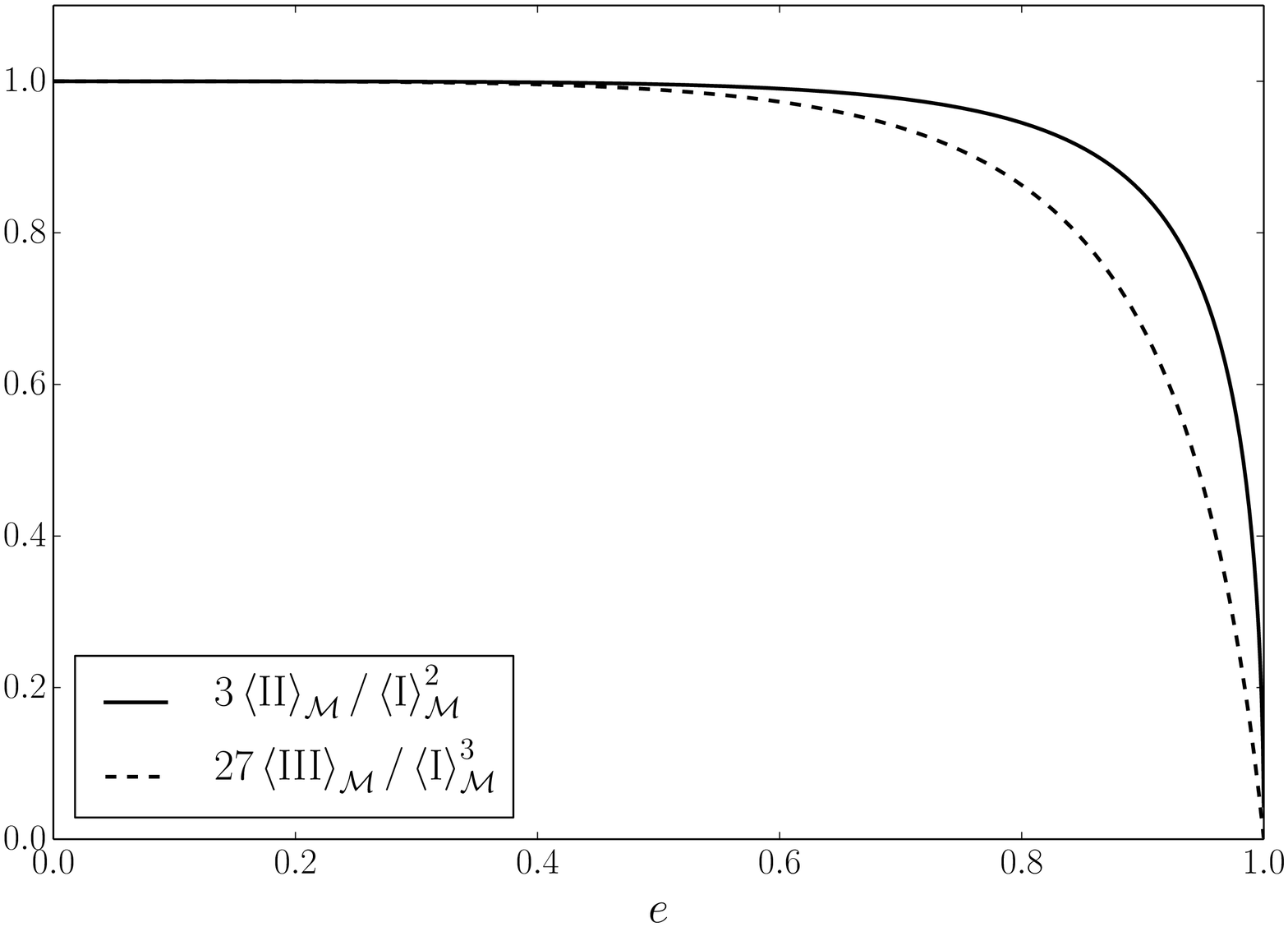}
\caption{Ratios $3\average{\inII}{M}/\average{\inI}{M}^2$ and $27\average{\inIII}{M}/\average{\inI}{M}^3$ as a function of the eccentricity of the oblate spheroid representing the domain $\CM$. The ratios do not depend on the major semi-axis of the spheroid. Thus, they do not depend on the size of the domain, which varies during the collapse (it increases before the turn-around, and then decreases until shell-crossing).}
\label{fig::ecc}
\end{figure}

Initially, the backreaction is close to zero, because of the quasi-sphericity of the domain, implying $\laverage{\inII}{\CM} \approx \laverage{\inI}{\CM}^2/3$ and $\laverage{\inIII}{\CM} \approx \laverage{\inI}{\CM}^3/27$. However, during the collapse, due to the increase of the eccentricity, $\average{\inII}{\CM}$, respectively $\average{\inIII}{\CM}$, becomes negligible compared with $\average{\inI}{\CM}^2$, respectively $\average{\inI}{\CM}^3$. This is depicted in figure~\ref{fig::ecc}, which shows the ratios $3\average{\inII}{\CM}/\average{\inI}{\CM}^2$ and $27\average{\inII}{\CM}/\average{\inI}{\CM}^3$ as a function of the eccentricity. This highlights the fact that the first invariant is the dominant term in the dynamical evolution of the domain $\CM$ in the phase before pancake formation. Thus, we can neglect the influence of the second and third invariants around the turn-around and at the end of the collapse of $\CM$. However, this is not necessarily valid at early stages of the collapse. In the next subsection we study the initial influence of the averaged second and third invariants on the subsequent evolution of the overdense domain, using the Relativistic Zel'dovich Approximation (RZA).

\subsection{Initial influence of the second and third averaged invariants}

We use the RZA equations presented in section~\ref{sec::TheRZA} with an EdS background. In equations~\eref{eq::RZA_QD} and \eref{eq::RZA_WD}, the terms featuring $\xi (t)$ are initially zero ($\initial{\xi} = 0$) but become non-negligible with increasing background scale factor $a(t)$, starting to impact around the turn-around time, and dominating during the final phase of the collapse. Thus, those terms are the ones which drive the end of the collapse. As we saw in the previous subsection for a collapsing oblate spheroid, the second and third invariants can be neglected compared with the first one during this phase, i.e. before pancake formation. We therefore assume $\laverage{\inII}{M} = 0$ and $\laverage{\inIII}{M} = 0$ in $\gamma_2$, $\gamma_3$,  $\tilde{\gamma}_2$ and $\tilde{\gamma}_3$ appearing in the equations~\eref{eq::RZA_QD} and \eref{eq::RZA_WD}. We are left with:
\begin{eqnarray}
\fl	\CQ_\CM = \frac{-2\initial{H}^2}{3a\left(a - 3\cstEdS\right)^2} \,  \left( 1 - 3 \frac{\laverage{\inII}{\CM}}{\laverage{\inI}{\CM}^2} \right) \ ; \
	\CW_\CM = \frac{-10\initial{H}^2}{a^2\left(a-3\cstEdS\right)} \,  \left( 1 + \frac{a \, \frac{\laverage{\inII}{\CM}}{\laverage{\inI}{\CM}^2}}{5\left(1-3\cstEdS\right)}\right) , \label{eq::EdS_AN}
\end{eqnarray}
where $\cstEdS$ is defined in section~\ref{sec::RZA_EdS}. Compared to the solutions found in section~\ref{sec::RZA_EdS}, $\CQ_\CM$ and $\CW_\CM$ have extra factors, which depend on $\laverage{\inII}{\CM}$. This highlights the fact that initially, for quasi-spherical domains, the second invariant and the first one squared are of the same order. Equation~\eref{eq::EdS_AN} leads to:
\begin{equation}
\fl	\uEdS = \frac{5/3}{a^2\left(a - 3\cstEdS\right)} \left( 1 + \frac{3 \, a \left(1-C\right)}{5\left(1-3\cstEdS\right)}\right) \quad ; \quad
	\partial_a \psiEdS = \pm\sqrt{\frac{2}{\kappa}}\sqrt{\frac{\epsilon\left(2a C-5\cstEdS\right)}{a\left(a-3\cstEdS\right)^2}} \, ,
	\label{eq::da_psi_100_AN}
\end{equation}
with $C := 1 - \laverage{\inII}{\CM} / 3\laverage{\inI}{\CM}^2$. As explained in the previous subsection, we assume the domain $\CM$ to be initially quasi-spherically symmetric. In a Newtonian approach, this implies $\laverage{\inII}{\CM} \approx \laverage{\inI}{\CM}^2/3$; we obtain $C \simeq 0.9$. As we are interested in collapsing domains, the initial shape of $\CM$ must be a pre-collapsed shape, i.e. an oblate spheroid with eccentricity $\gtrsim$ 0. Then, equations~\eref{eq::W0_AN} to \eref{eq::I_II_AN} imply $\laverage{\inII}{\CM} \lesssim \laverage{\inI}{\CM}^2/3$. So, $0.9 \lesssim C \leqslant 1$. In this range, the behaviour of the morphon potential remains the same as in section~\ref{sec::RZA_EdS}, only the timing changes: the maximum of the potential and the transition of the field between a phantom field to a real field are reached at a later time. The amplitude of the curvature is of the same order as in section~\ref{sec::TheRZA}, hence the results of this subsection and the following ones hold, even if we take a non-zero second initial invariant which is representative of a pre-collapsing overdense domain. This justifies the hypothesis chosen all along this article to take $\laverage{\inI}{\CM} \not= 0$, $\laverage{\inII}{\CM} = 0$ and $\laverage{\inIII}{\CM} = 0$, for an oblate spheroid overdense domain.

Some further supporting aspects of this idealization compared with a generic realization of RZA can be found in \cite{JanBoud}.

\section{Turn-around fit between RZA and scaling solution models}
\label{appB}

In order to reduce the set of free parameters in the scaling solution model with respect to the RZA-EdS model, we have assumed $t_{\rm ta}^{\rm SS} = t_{\rm ta}^{\rm RZA}\left(\cstEdS, \initial H\right)$ and $a_{\CM_{\rm ta}}^{\rm SS} = a_{\CM_{\rm ta}}^{\rm RZA}\left(\cstEdS\right)$. \\
We have
\begin{eqnarray}
	\fl t_{\rm ta}^{\rm RZA}\left(\cstEdS, \initial H\right) = \frac{2}{3\initial H} \left[ \left(\frac{9\cstEdS}{4}\right)^{3/2} - 1 \right] \quad ; \quad a_{\CM_{\rm ta}}^{\rm RZA}\left(\cstEdS\right) = \frac{9\cstEdS}{4} \left(\frac{3\cstEdS/4}{3\cstEdS - 1}\right)^{1/3} \, , \label{eq::ta_RZA}
\end{eqnarray}
for the RZA turn-around. For the scaling solution, the domain scale factor at turn-around is given by $a_{\CM_{\rm ta}}^{\rm SS} = \left(-\gamma^{\initial\CM}_{\CW\CQ\mathrm{m}}\right)^{\frac{-1}{n+3}}$. Using $a_{\CM_{\rm ta}}^{\rm SS} = a_{\CM_{\rm ta}}^{\rm RZA}\left(\cstEdS\right)$, this gives:
\begin{eqnarray}
	-\gamma^{\initial\CM}_{\CW\CQ\mathrm{m}} = \left( \frac{9\cstEdS}{4} \left(\frac{3\cstEdS/4}{3\cstEdS - 1}\right)^{1/3} \right)^{-(n+3)} \, . \label{eq::gamma_SS_RZA_bis}
\end{eqnarray}
From equations~\eref{eq::averaged}, with $\Lambda = 0$ and the scaling solution ansatz, we can derive the time $t^{\rm SS}$ as a function of the domain scale factor $a^{\rm SS}_\CM$. We have
\begin{eqnarray}
	\fl t^{\rm SS}\left(A_\CM\right) = \frac{2}{3H^{\rm SS}_\initial{\CM}} \sqrt{\frac{1+\gamma^{\initial\CM}_{\CW\CQ\mathrm{m}}}{\left(-\gamma^{\initial\CM}_{\CW\CQ\mathrm{m}}\right)^{3/(n+3)}}} \, \left\{ \CB_n\left(A_\CM\right) - \CB_n\left(\left[-\gamma^{\initial\CM}_{\CW\CQ\mathrm{m}}\right]^{\frac{1}{n+3}}\right) \right\} \, , \label{eq::ta_SS}
\end{eqnarray}
with $ \CB_n\left(A_\CM\right) := A_\CM^{3/2} \ {_2}F_1\left(\frac{1}{2}, \frac{3/2}{n+3}; 1+ \frac{3/2}{n+3}; A_\CM^{3+n}\right)$ and $A_\CM := a_\CM \, \left[-\gamma^{\initial\CM}_{\CW\CQ\mathrm{m}}\right]^{\frac{1}{n+3}}$.

\begin{figure}[h]
	\centering
	\includegraphics[width=0.6\textwidth]{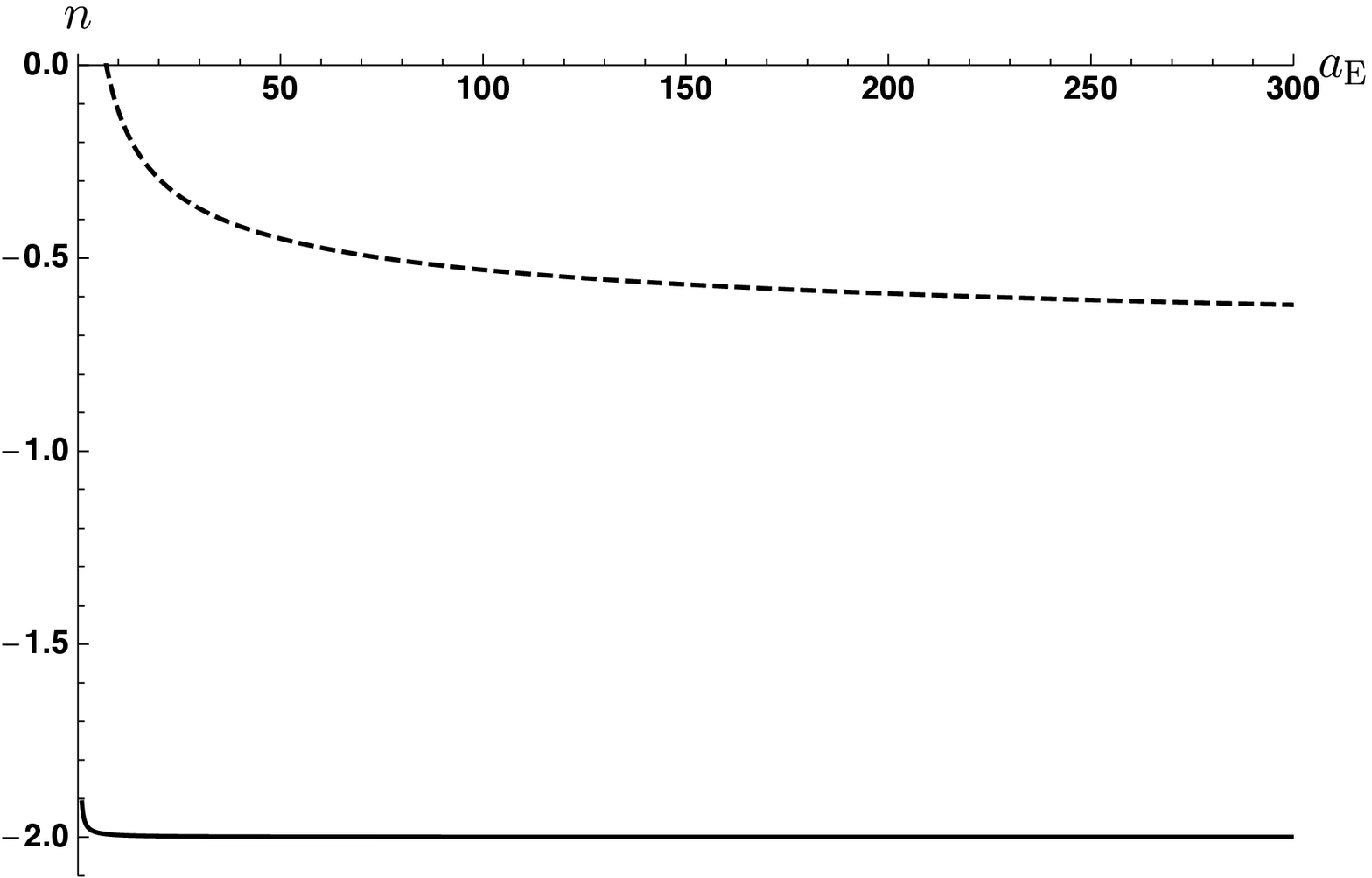}
	\caption{The scaling solutions $n_{-2}$ (full line) and $n_{-1}$ (dashed line)  as a function of $\cstEdS$ for realistic values of $\cstEdS$ given in table~\ref{tab::mass}.}
	\label{fig::soln_SS}
\end{figure}

Then, using equations~\eref{eq::gamma_SS_RZA}, \eref{eq::ta_RZA}, \eref{eq::gamma_SS_RZA_bis} and \eref{eq::ta_SS}, the exponent $n$ is the solution of the following equation:
\begin{eqnarray}
\fl \frac{9}{4}\cstEdS^{3/2} - \frac{2}{3} = \left(3\cstEdS-1\right)\sqrt{n+2} \left[-\gamma^{\initial\CM}_{\CW\CQ\mathrm{m}}\right]^{\frac{n/2}{n+3}} \left\{ \CB_n\left(1\right) - \CB_n\left(\left[-\gamma^{\initial\CM}_{\CW\CQ\mathrm{m}}\right]^{\frac{1}{n+3}}\right) \right\} \, , \label{eq::soln_SS}
\end{eqnarray}
and the parameters $\gamma^{\initial\CM}_{\CW\CQ\mathrm{m}}$ and $H^{\rm SS}_\initial{\CM}$ are given by:
\begin{eqnarray}
\fl
	\gamma^{\initial\CM}_{\CW\CQ\mathrm{m}} = -\left( \frac{9\cstEdS}{4} \left(\frac{3\cstEdS/4}{3\cstEdS - 1}\right)^{1/3} \right)^{-(n+3)} \ ; \ 
	H^{\rm SS}_\initial{\CM} = \frac{2\initial H}{3\left(3\cstEdS-1\right)}\sqrt{\frac{1+\gamma^{\initial\CM}_{\CW\CQ\mathrm{m}}}{-\gamma^{\initial\CM}_{\CW\CQ\mathrm{m}}\left(n+2\right)}} \ \ .
\end{eqnarray}
Recalling that $n > -2$, equation~\eref{eq::soln_SS} has two solutions for $n\left(\cstEdS\right)$. We denote them by $n_{-2}$ and $n_{-1}$. They are shown in figure~\ref{fig::soln_SS} for $\cstEdS \in [\frac{1}{3} ; 300]$. The solution $n_{-2}$ converges (from above) quickly to $-2$ in the limit $\cstEdS \rightarrow +\infty$, such that for realistic values of $\cstEdS$, given in table~\ref{tab::mass}, we have $n_{-2} \simeq -2$. The solution $n_{-1}$ converges (from above) very slowly to $-1$. In the range of realistic values for $\cstEdS$, we have $n_{-1} \simeq -0.5$.

\begin{figure}
	\centering
	\includegraphics[width=0.6\textwidth]{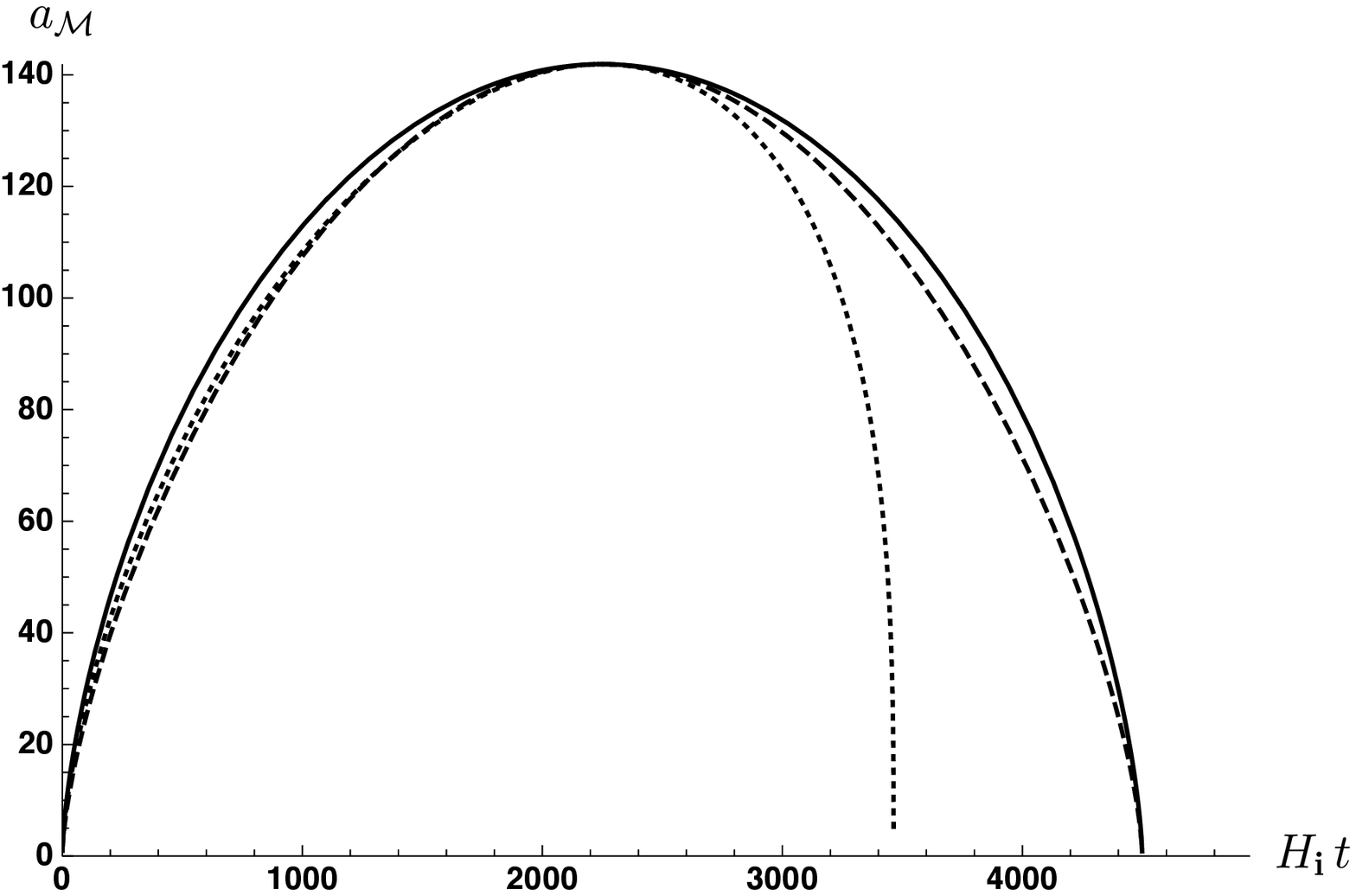}
	\caption{Evolution of the domain scale factor $a_\CM$ as a function of $\initial H \, t$ for the RZA-EdS model (short-dashed line), the $n_{-2}$ scaling solution (full line), and the $n_{-1}$ scaling solution (long-dashed line), for $\cstEdS = 100$.}
	\label{fig::aM_SS}
\end{figure}

In figure~\ref{fig::aM_SS}, we show, for $\cstEdS = 100$, the evolution of the domain scale factor $a_\CM$ as a function of the adimensional time $\initial H \, t$ for the RZA-EdS model, the $n_{-2}$ scaling solution, and the $n_{-1}$ scaling solution. The first phase of the collapse, i.e. until the turn-around, is similar for the three models. However, the scaling solutions diverge from the RZA-EdS solution just after the turn-around, which leads to a later collapse. This shows that scaling solutions are not realistic for the whole period of collapse due to their time-symmetric behaviour at 
turn-around.

\bigskip\bigskip
\section*{References}
\providecommand{\newblock}{}

\end{document}